%% file: paper.tex
\documentclass[letterpaper,twocolumn,10pt]{article}
\usepackage[twoside=true, head=13pt,
     paperwidth=8.5in, paperheight=11in,
     includeheadfoot=false, columnsep=2pc,
     top=1in, bottom=1in, inner=0.75in, outer=0.75in,
     marginparwidth=2pc,heightrounded]{geometry}

\usepackage{graphicx}
\usepackage[tt=false, type1=true]{libertine}
\usepackage{mathptmx}
\usepackage{amssymb}
\usepackage[varqu]{zi4}
\usepackage[T1]{fontenc}
\usepackage{enumitem}
\usepackage[font={footnotesize},labelfont={footnotesize,bf},textfont={footnotesize,it}]{caption}
\usepackage{lipsum}

\usepackage{xcolor}
\usepackage{tcolorbox}
\tcbuselibrary{breakable}
\usepackage{amsmath}
\usepackage{booktabs}
\usepackage{multirow}
\usepackage{subcaption}   %
\usepackage{graphicx}     %
\usepackage{bm}
\usepackage{algorithm}
\usepackage{algpseudocode}
\usepackage[]{hyperref}         
\hypersetup{                    
  colorlinks,
  linkcolor={green!80!black},
  citecolor={red!70!black},
  urlcolor={blue!70!black}
}
\usepackage{tabularx}
\usepackage[table]{xcolor} 

\title{\large{\it CCA Reimagined:}\\
\Large \bf An Exploratory Study of Large Language Models for Congestion Control}
\author{
Xiaoxuan Qin$^{*}$ \quad Yufei Wang$^{*}$ \quad Longfei Shangguan \\
School of Information and Computing \\
University of Pittsburgh, Pittsburgh, USA \\
\texttt{\{xiq33, yuw384, longfei\}@pitt.edu}
}

\date{}

\begin{document}
\pagestyle{plain}
\maketitle
\renewcommand{\thefootnote}{\fnsymbol{footnote}}
\footnotetext[1]{These authors contributed equally as first authors.}
\begin{abstract}
In this paper, we conduct an emulation-guided study to systematically investigate the feasibility of Large language model (LLM)-driven congestion control.
The exploration is structured into two phases. The first phase derisks the whole capability where we isolate the role of LLM on a single yet crucial congestion avoidance phase so that we can safely examine when to invoke the LLM, what information to provide, and how to formulate LLM instructions.
Based on the gained insights, we extend LLM's role to multiple congestion control phase and propose a more generic LLM-based congestion control policy. Our evaluation on both static and dynamic network traces demonstrates that the LLM-based solution can reduce latency by up to 50\% with only marginal throughput sacrifice (e.g., less than 0.3\%) compared to traditional CCAs.
Overall, our exploration study confirms the potential of LLMs for adaptive and general congestion control, demonstrating that when granted appropriate control freedom and paired with an effective triggering mechanism, LLM-based policies achieve significant performance gains, particularly under highly dynamic network conditions.

\end{abstract}


\input{1intro}

\input{2related}

\input{3overview}

\input{4limited}

\input{5free}

\input{6fairness}
\input{7interpretability}

\input{8conclusion}

\newpage
\bibliographystyle{plain}
\bibliography{sample}
\clearpage
\appendix

\section{Prompts}
In this section, we present all three types of prompts we used in our experiments. \\
\begin{tcolorbox}[
    colframe=blue!50!black,
    colback=white,
    coltitle=white,
    fonttitle=\bfseries,
    title=TCP-LLM-L Prompt,
    boxrule=1.2pt,
    width=\textwidth
    ]
{\footnotesize
\vspace{-2mm}
\textbf{System Prompt:}\\
You are an expert congestion controller in wide area networks. 
Based on the following input parameters: 
1. last\_cwnd: The previous congestion window size.
2. current\_cwnd: The current size of the congestion window.
3. ssthreshold: The slow start threshold.
4. last\_rtt: The previous round-trip time of second-to-last segment.
5. current\_rtt: The current round-trip time of last segment.
6. current\_throughput: The current throughput observed (in Mbps) during the latest sampling interval.
7. last\_throughput: The throughput observed (in Mbps) during the previous sampling interval.

Explanation of the history fields:

- CWNDs (most recent first):
  A list of recently recorded congestion window sizes (in bytes). Each value corresponds to the cwnd measured at a discrete time step, with the most recent measurement appearing first.

- RTTs (most recent first):
  A list of recently recorded round-trip times (in seconds). Each value is the measured latency for a packet to travel from sender to receiver and back again, with the most recent measurement first.

- Throughput values (most recent first):
  A list of recently recorded throughput measurements (in Mbps). Each value represents the data transfer rate observed during a sampling interval, listed from the newest to the oldest measurement.

Modify cwnd and ssthreshold according to the following rules: \\
1. If current\_rtt is higher than last\_rtt (indicating rising latency) and current\_throughput shows no significant improvement over or is equal to or lower than last\_throughput (indicating congestion and buffer buildup), reduce cwnd multiplicatively, but limit the reduction to no more than 50\% of the current\_cwnd.\\
2. If current\_cwnd is stable or slightly lower than last\_cwnd, and current\_rtt remains stable at a low level (or shows a decreasing trend) while current\_throughput is increasing compared to last\_throughput, increase cwnd additively(e.g., cwnd += 1448), as it's likely that the network can handle more traffic. \\
3. Do not allow cwnd to fall below 1448 bytes, which is one segment size. In addition, prevent any single adjustment from causing cwnd to drop abruptly to 1448.\\
4. Limit the rate of change in cwnd to avoid oscillations (e.g., no more than ±50\% per adjustment).\\
5. Do not increase cwnd if current\_rtt is rising rapidly and current\_throughput is not showing signigficantly improvment compared to last\_throughput. \\
6. Upon detecting congestion (e.g., packet loss or a RTT increase), dynamically set the ssthreshold as a proportion of the current cwnd:
   - In severe congestion (when current\_throughput is stagnant), set ssthreshold to the current cwnd.
   - In favorable conditions (when current\_throughput is increasing), allow ssthreshold to reach up to 1.5 (or 1.25) times the current cwnd.
   Use a smoothing mechanism to ensure gradual transitions in ssthreshold.\\
7. If, after cwnd has been reduced, current\_throughput remains stagnant (i.e., shows no significant improvement) and current\_rtt shows little decline or same compared to last\_rtt (indicating persistent buffering), further reduce cwnd multiplicatively (using the same method as in Rule 1) until a measurable drop in RTT is achieved.\\
Suggest the optimal congestion control parameters for the next step in the json format with only values without any explanation. You should generate the following parameters:\\
"next\_CWND"\\
"next\_SSThreshold"

\vspace{-2mm}
}
\end{tcolorbox}
\vspace{-4mm}
\captionsetup{type=table}
\captionof{table}{Prompt utilized in TCP-LLM baseline design.}
\label{tab:gpt_prompts_baseline}

\begin{table*}[t]{
\begin{tcolorbox}[
    colframe=blue!50!black,
    colback=white,
    coltitle=white,
    fonttitle=\bfseries,
    title=TCP-LLM-G Prompt,
    boxrule=1.2pt
    ]

{\footnotesize
\vspace{-2mm}
\textbf{System Prompt:}\\
You are an expert congestion controller in wide area networks. Based on the following input parameters:
1. last\_cwnd: CWND measured at the previous interval (bytes).
2. current\_cwnd: CWND measured at the current interval (bytes).
3. ssthreshold: Current slow start threshold (bytes).
4. last\_rtt: RTT observed in the previous interval (seconds).
5. current\_rtt: RTT observed in the current interval (seconds).
6. last\_throughput: Throughput in the previous interval (Mbps).
7. current\_throughput: Throughput in the current interval (Mbps).
8. current\_retransmit\_packet: Number of retransmitted packets in the current interval.
9. history\_cwnd: List of the last N CWND measurements (bytes), most recent first.
10. history\_rtt: List of the last N RTT measurements (s), most recent first.
11. history\_throughput: List of the last N throughput values (Mbps), most recent first.
12. history\_retransmit\_packet: List of the last N retransmit packet counts, most recent first.

Define adjustments to cwnd and ssthreshold according to these principles:\\
1. Decrease cwnd multiplicatively if there might be a congestion.\\
2. Increase cwnd additively if there might be underutilization of the bandwidth.\\
3. A congestion might be happening if you see current\_rtt is higher than the most recent value of history\_rtt, while current\_throughput or history\_throughput is not in an increasing trend or it is even in a decreasing trend.\\
4. A underutilization of the bandwidth might be happening if you see current\_cwnd and history\_cwnd are decreasing and current\_throughput and history\_throughput are decreasing while a decreasing trend in current\_rtt or history\_rtt.\\
5. If current\_rtt or history\_rtt remain elevated compared to recent baselines, and current\_throughput or history\_throughput are flat or degrading (not showing meaningful improvement), treat this as queue buildup. Immediately decrease cwnd by a noticeable fraction of its current size to lower RTT and stabilize throughput.\\
6. Do not allow cwnd to fall below 1448 bytes, which is one segment size.\\
7. Avoid any single adjustment from causing cwnd to drop abruptly to 1448.\\
8. Upon detecting congestion (e.g.,  packet retransmit or a RTT increase), set ssthreshold to the new cwnd value you suggested.\\
9. Ideally, we want the current retransmit packet to be 0. If the current retransmit packet is higher than 0, you should immediately reduce cwnd multiplicatively by a large fraction of its current size. \\
10. When you believe this is a dynamic network not a static network, conduct this rule when rule 9 is not satisfied and rule 5 is not satisfied. If current\_cwnd and history\_cwnd keep stable or increasing while history\_throughput is in an increasing trend and current\_throughput is significantly improved, and the trend of current\_rtt and history\_rtt are in a decreasing trend, immediately increase cwnd additively by a substantial fraction of its current size that reflects the scale of throughput improvement, keep the increase only if throughput immediately improves, otherwise revert.\\
11. If you observe your cwnd is not decreasing while RTT is increasing, treat this as a sign that there are other nodes in the network. Try to maintain fairness across all senders. Do not back off too much.\\
12. If you observe your current\_cwnd and history\_cwnd remain less than the inital cwnd 14480, 10 segments, an extremely low level, you should immediately recover the cwnd to at least 14480(10 segments) aggresively to catch up with others and maintain fairness, ignore the current\_retransmit\_packet.\\
Suggest the optimal congestion control parameters for the next step in the json format with only values without any explanation. You should generate the following parameters:\\
"next\_CWND"\\
"next\_SSThreshold"
\vspace{-2mm}
}
\end{tcolorbox}
}
\vspace{-4mm}
\caption{Prompt utilized in TCP-LLM generalized design.}
\label{tab:gpt_prompts_generalized}
\end{table*}

\clearpage
\begin{table*}[t]{
\begin{tcolorbox}[
    colframe=blue!50!black,
    colback=white,
    coltitle=white,
    fonttitle=\bfseries,
    title=TCP-LLM-G Prompt with an Additional Aggressive Rule,
    boxrule=1.2pt
    ]

{\footnotesize
\vspace{-2mm}
\textbf{System Prompt:}\\
You are an expert congestion controller in wide area networks. Based on the following input parameters:
1. last\_cwnd: CWND measured at the previous interval (bytes).
2. current\_cwnd: CWND measured at the current interval (bytes).
3. ssthreshold: Current slow start threshold (bytes).
4. last\_rtt: RTT observed in the previous interval (seconds).
5. current\_rtt: RTT observed in the current interval (seconds).
6. last\_throughput: Throughput in the previous interval (Mbps).
7. current\_throughput: Throughput in the current interval (Mbps).
8. current\_retransmit\_packet: Number of retransmitted packets in the current interval.
9. history\_cwnd: List of the last N CWND measurements (bytes), most recent first.
10. history\_rtt: List of the last N RTT measurements (s), most recent first.
11. history\_throughput: List of the last N throughput values (Mbps), most recent first.
12. history\_retransmit\_packet: List of the last N retransmit packet counts, most recent first.

Define adjustments to cwnd and ssthreshold according to these principles:\\
1. Decrease cwnd multiplicatively if there might be a congestion. \\
2. Increase cwnd additively if there might be underutilization of the bandwidth. \\
3. A congestion might be happening if you see current\_rtt is higher than the most recent value of history\_rtt, while current\_throughput or history\_throughput is not in an increasing trend or it is even in a decreasing trend.\\
4. A underutilization of the bandwidth might be happening if you see current\_cwnd and history\_cwnd are decreasing and current\_throughput and history\_throughput are decreasing while a decreasing trend in current\_rtt or history\_rtt.\\
5. If current\_rtt or history\_rtt remain elevated compared to recent baselines, and current\_throughput or history\_throughput are flat or degrading (not showing meaningful improvement), treat this as queue buildup. Immediately decrease cwnd by a noticeable fraction of its current size to lower RTT and stabilize throughput.\\
6. Do not allow cwnd to fall below 1448 bytes, which is one segment size.\\
7. Avoid any single adjustment from causing cwnd to drop abruptly to 1448.\\
8. Upon detecting congestion (e.g.,  packet retransmit or a RTT increase), set ssthreshold to the new cwnd value you suggested.\\
9. Ideally, we want the current retransmit packet to be 0. If the current retransmit packet is higher than 0, you should immediately reduce cwnd multiplicatively by a large fraction of its current size. \\
10. When you believe this is a dynamic network not a static network, conduct this rule when rule 9 is not satisfied and rule 5 is not satisfied. If current\_cwnd and history\_cwnd keep stable or increasing while history\_throughput is in an increasing trend and current\_throughput is significantly improved, and the trend of current\_rtt and history\_rtt are in a decreasing trend, immediately increase cwnd additively by a substantial fraction of its current size that reflects the scale of throughput improvement, keep the increase only if throughput immediately improves, otherwise revert.\\
11. If you observe your cwnd is not decreasing while RTT is increasing, treat this as a sign that there are other nodes in the network. Try to maintain fairness across all senders. Do not back off too much.\\
12. If you observe your current\_cwnd and history\_cwnd remain less than the inital cwnd 14480, 10 segments, an extremely low level, you should immediately recover the cwnd to at least 14480(10 segments) aggresively to catch up with others and maintain fairness, ignore the current\_retransmit\_packet.\\
\textbf{13. If you think the current throughput is low and current cwnd is low. Try to spike up like Bbr to probe the bandwidth.\\}
Suggest the optimal congestion control parameters for the next step in the json format with only values without any explanation. You should generate the following parameters:\\
"next\_CWND"\\
"next\_SSThreshold"
\vspace{-2mm}
}
\end{tcolorbox}
}
\vspace{-4mm}
\caption{TCP-LLM-G Prompt with an Additional Aggressive Rule}
\label{tab:fairness_prompt_scheme}
\end{table*}
\clearpage
\section{Trigger Threshold Ablation Study}

\begin{table}[h]
  \centering
  \begin{tabular}{lccccc}
    \toprule
    \multirow{2}{*}{$\alpha$} & \multicolumn{2}{c}{RTT (ms)} & \multicolumn{2}{c}{Thr. (Mbps)} \\
    \cmidrule(lr){2-3}\cmidrule(lr){4-5}
     & Avg. & Std. & Avg. & Std. \\
    \midrule
    50\% & 75.62  & 7.18 & 9.98 & 0.00 \\
    60\% & 99.55 & 7.24 & 9.99 & 0.00 \\
    70\% & \textbf{92.83}  & 3.54 & \textbf{9.99} & 0.00 \\
    80\% & 115.40 & 3.43 & 9.99 & 0.00 \\
    \bottomrule
  \end{tabular}
  \caption{Trigger threshold $\alpha$ ablation.}
  \vspace{-4mm}
  \label{tab:alpha_ablation}
\end{table}

\subsection{$\alpha$ in latency-based trigger model}

\label{sec:latency_trigger_ablation}
Our current design relies on a pre-defined threshold, the natural question raises is whether the system performance is sensitive to this threshold. We conduct experiments on the choice of this threshold $\alpha$ ranging from 50\% to 90\% and test the RTT and throughput in each settings.  The results are shown in Tab.~\ref{tab:alpha_ablation}.
We observe a clear tradeoff driven by the trigger threshold $\alpha$. Lower thresholds (e.g., 50\%) invoke the LLM more aggressively, which helps keep latency low but slightly sacrifices throughput due to frequent adjustments that interrupt stable growth. Conversely, higher thresholds (e.g., 80\%) reduce responsiveness between decisions, which increases throughput stability but at the cost of building longer queues, leading to higher latency. The intermediate setting of $\alpha = 70\%$ achieves the best balance. It avoids excessive oscillations while still reacting quickly enough to prevent significant queue buildup, thereby sustaining near-maximal throughput with improved RTT performance. It is worth noticing that for all thresholds, our method has a lower latency than the baseline TCP-NewReno. 

\begin{table}[h]
  \centering
  \begin{tabular}{lccccc}
    \toprule
    \multirow{2}{*}{$\beta$} & \multicolumn{2}{c}{RTT (ms)} & \multicolumn{2}{c}{Thr. (Mbps)} \\
    \cmidrule(lr){2-3}\cmidrule(lr){4-5}
     & Avg. & Std. & Avg. & Std. \\
    \midrule
    0.05 & 97.72  & 8.80 & 9.98 & 0.00 \\
    0.1 & \textbf{91.32} & 0.16 & \textbf{9.99} & 0.00 \\
    0.15 & 91.39  & 0.35 & 9.99 & 0.00 \\
    0.2 & 92.74 & 2.43 & 9.99 & 0.00 \\
    \bottomrule
  \end{tabular}
  \caption{Trigger threshold $\beta$ ablation.}
  \vspace{-4mm}
  \label{tab:beta_ablation}
\end{table}

\subsection{$\beta$ in ACK-based trigger model}
\label{ss:impact_of_beta}

We select the range of $\beta$ between 0.05 and 0.2 based on preliminary experiments with TCP-NewReno, where roughly $8{,}000$ ACKs are received within 10 seconds in the static trace. The rationale is straightforward: if $\beta$ is set too small, the ACK-based trigger would invoke the LLM excessively, incurring unnecessary overhead and cost; if $\beta$ is set too large, the trigger would become too sparse, potentially missing important network dynamics. As reported in Table~\ref{tab:beta_ablation}, our ablation results demonstrate that the choice of $\beta$ has only a marginal impact on the overall performance of TCP-LLM-G. The main consideration, therefore, lies in balancing responsiveness with practicality. To achieve this balance, we adopt $\beta = 0.1$, corresponding to an ACK threshold of $800$, as the default configuration in our subsequent experiments.
\section{Fairness}

\begin{table}[h]
\centering
\resizebox{\columnwidth}{!}{%
\begin{tabular}{lccccc}
\toprule
\textbf{Node} & \textbf{CCA} & \textbf{Static} & \textbf{LongIsland} & \textbf{7Train} & \textbf{QTrain} \\
\midrule
\multicolumn{6}{l}{\textbf{TCP-LLM-L}} \\
Node 0 & LLM & 35.60\% & 28.60\% & 33.20\% & 31.30\% \\
Node 1 & LLM & 29.60\% & 32.30\% & 27.60\% & 33.80\% \\
Node 2 & LLM & 33.20\% & 31.70\% & 33.30\% & 33.10\% \\
\textbf{SUM} & & \textbf{98.40\%} & \textbf{92.60\%} & \textbf{94.10\%} & \textbf{98.20\%} \\
\midrule
\multicolumn{6}{l}{\textbf{TCP-LLM-G - ALL LLM}} \\
Node 0 & LLM & 26.10\% & 33.40\% & 33.00\% & 42.30\% \\
Node 1 & LLM & 36.10\% & 35.00\% & 29.30\% & 27.30\% \\
Node 2 & LLM & 37.30\% & 27.60\% & 33.90\% & 28.70\% \\
\textbf{SUM} & & \textbf{99.50\%} & \textbf{96.00\%} & \textbf{96.20\%} & \textbf{98.30\%} \\
\midrule
\multicolumn{6}{l}{\textbf{TCP-LLM-G - Hybrid}} \\
Node 0 & LLM & 24.10\% & 9.10\% & 14.80\% & 21.50\% \\
Node 1 & LLM & 27.20\% & 39.10\% & 20.70\% & 11.20\% \\
Node 2 & Bbr & 48.00\% & 48.10\% & 60.50\% & 66.20\% \\
\textbf{SUM} & & \textbf{99.30\%} & \textbf{96.30\%} & \textbf{96.00\%} & \textbf{99.00\%} \\
\bottomrule
\end{tabular}%
}
\caption{Fairness comparison: Bandwidth utilization of different nodes.}
\label{tab:fairness_table}
\end{table}

To evaluate fairness in the all-LLM setting, we compute the bandwidth utilization of each flow at one-second granularity and then average these values over the entire 120 s trace. This produces an average bandwidth utilization per flow, as summarized in Table~\ref{tab:fairness_table}. The results show that both TCP-LLM-L and TCP-LLM-G exhibit relatively balanced allocations: across all four topologies, the per-flow averages differ by less than 5\%. This indicates that on long timescales, our LLM-based schemes maintain a high degree of fairness.
\clearpage
\section{LLM generated algorithm and explanation}
\begin{algorithm}[h]
\caption{LLM-Inspired Congestion Control (from QTrain run)}
\label{alg:llm_heuristic}
\begin{algorithmic}[1]
\State \textbf{Params:} $\beta_{\text{heavy}}{=}0.5$, $\beta_{\text{mild}}{=}0.75$, 
$\gamma_{\text{cong}}{=}0.9$, $\Delta$, $T_{\text{probe}}$, 
$\varepsilon_{rtt}$, $\varepsilon_{rtt}^{+}$, $R_{\text{heavy}}$, \texttt{wait}=2s
\vspace{2pt}
\If{\texttt{loss\_triggered} (\,$retx_t > 0$\,)} \\ 
\Comment{1) Loss handling: Two-step multiplicative cut.}
    \If{$retx_t \ge R_{\text{heavy}}$ \textbf{or} \texttt{LOSS} high}
        \State $cwnd \gets \max(\lfloor \beta_{\text{heavy}}\!\cdot\! cwnd \rfloor,\, 1\!\cdot\! MSS)$
    \Else
        \State $cwnd \gets \max(\lfloor \beta_{\text{mild}}\!\cdot\! cwnd \rfloor,\, 1\!\cdot\! MSS)$
    \EndIf
    \State $ssthresh \gets cwnd$
    \State \texttt{arm\_wait\_timer}(\texttt{2s})

\ElsIf{\texttt{congestion\_gate} (\,$retx_t{=}0$ \textbf{and} $dRTT \ge \varepsilon_{rtt}^{+}$\,)} \\
\Comment{2) Congestion Handling: Small cut ($\approx10\%$).}
    \State $cwnd \gets \max(\lfloor \gamma_{\text{cong}}\!\cdot\! cwnd \rfloor,\, 1\!\cdot\! MSS)$
    \State $ssthresh \gets cwnd$
    \State \texttt{arm\_wait\_timer}(\texttt{2s})

\ElsIf{\texttt{probe\_triggered} (\,$time\_since\_last\_action \ge T_{\text{probe}}$\,) \\ \textbf{and}\\
\hspace{2.2em}\texttt{stability\_gate} (\,$retx_t{=}0$ \textbf{and} $|dRTT| \le \varepsilon_{rtt}$\,) \textbf{and}\\
\hspace{2.2em}\texttt{throughput\_gate} (\,$dTP \ge 0$\,)}\\
\Comment{3) Probing bandwidth: Conservative additive.}
    \State $cwnd \gets cwnd + \Delta$
    \State $ssthresh \gets cwnd$
    \State \texttt{arm\_wait\_timer}(\texttt{2s})

\Else
\Comment{4) Keep hold: Rising RTT/ambiguous signals.}
    \State $cwnd \gets cwnd$
\EndIf
\end{algorithmic}
\end{algorithm}

\begin{tcolorbox}[
  breakable,
  colback=white, colframe=black, boxrule=0.6pt, arc=6pt,
  title=Rule Explanations (for Alg.~\ref{alg:llm_heuristic}),
  colbacktitle=black, coltitle=white, fonttitle=\bfseries,
  before upper=\setlength{\parskip}{4pt}\setlength{\parindent}{0pt}\raggedright
]

\textbf{Rule 1 — Loss Handling.}
(1) If retransmissions $>0$, reduce \texttt{CWND} immediately.
(2) For heavy loss (e.g., $\ge 3$ retransmits or persistent $\text{EWMA}$ loss), cut by $50\%$ ($\beta_{\text{heavy}}=0.5$);
for mild loss (1--2 retransmits), cut by $25\%$ ($\beta_{\text{mild}}=0.75$).
(3) Reset $ssthresh$ to the new $cwnd$ to avoid re-entering slow start.
This matches the sharp two-step reductions observed in the log (e.g., $322\mathrm{k}\!\to\!242\mathrm{k}\!\to\!121\mathrm{k}$).

\textbf{Rule 2 — Congestion Handling.}  
If RTT consistently increases beyond a small threshold ($|dRTT| > \varepsilon_{\text{rtt}}$) without retransmissions,  
interpret this as early congestion from queue buildup. In this case, reduce \texttt{CWND} moderately (e.g., by 10\%),  
allowing queues to drain before further probing.  
This prevents latency inflation even in the absence of packet loss.

\textbf{Rule 3 — Stability-Gated Probing.}
Every $T_{\text{probe}}$ ($\approx 1.5\text{--}2\,\text{s}$), if no retransmissions are observed, the RTT change is small ($|dRTT|\le \varepsilon_{\text{rtt}}$, about 2--3\,ms),
and throughput is stable or increasing ($dTP \ge 0$),
increase CWND by $\Delta$ (about $+1\text{--}3$\,MSS in normalized units).
This mirrors the cautious, stepwise increases observed after stability (e.g., $121632 \to 122632 \to \cdots$).

\textbf{Rule 4 — Holding.}
If neither loss nor probe conditions apply, hold CWND.
This corresponds to plateaus where throughput recovers and RTT stabilizes without overshoot.

\textbf{Suggested defaults by scenario.}
Static/low-variance: $T_{\text{probe}}=2\,\text{s}$, $\Delta=1\,\text{MSS}$, $\varepsilon_{\text{rtt}}=2\,\text{ms}$.
Moderate variability: $T_{\text{probe}}=1.5\,\text{s}$, $\Delta=2\,\text{MSS}$, $\varepsilon_{\text{rtt}}=3\,\text{ms}$.
Highly fluctuating (QTrain-like): $T_{\text{probe}}=1\text{--}1.5\,\text{s}$, $\Delta=2\text{--}3\,\text{MSS}$, $\varepsilon_{\text{rtt}}=4\text{--}5\,\text{ms}$, $R_{\text{heavy}}=3$.
Long-RTT (satellite): $T_{\text{probe}}=3\text{--}4\,\text{s}$, $\Delta=1\,\text{MSS}$, $\varepsilon_{\text{rtt}}=10\,\text{ms}$.
\end{tcolorbox}
\label{tab;gpt_explanation}

\begin{figure}[h]
    \centering
    \includegraphics[width=\columnwidth]{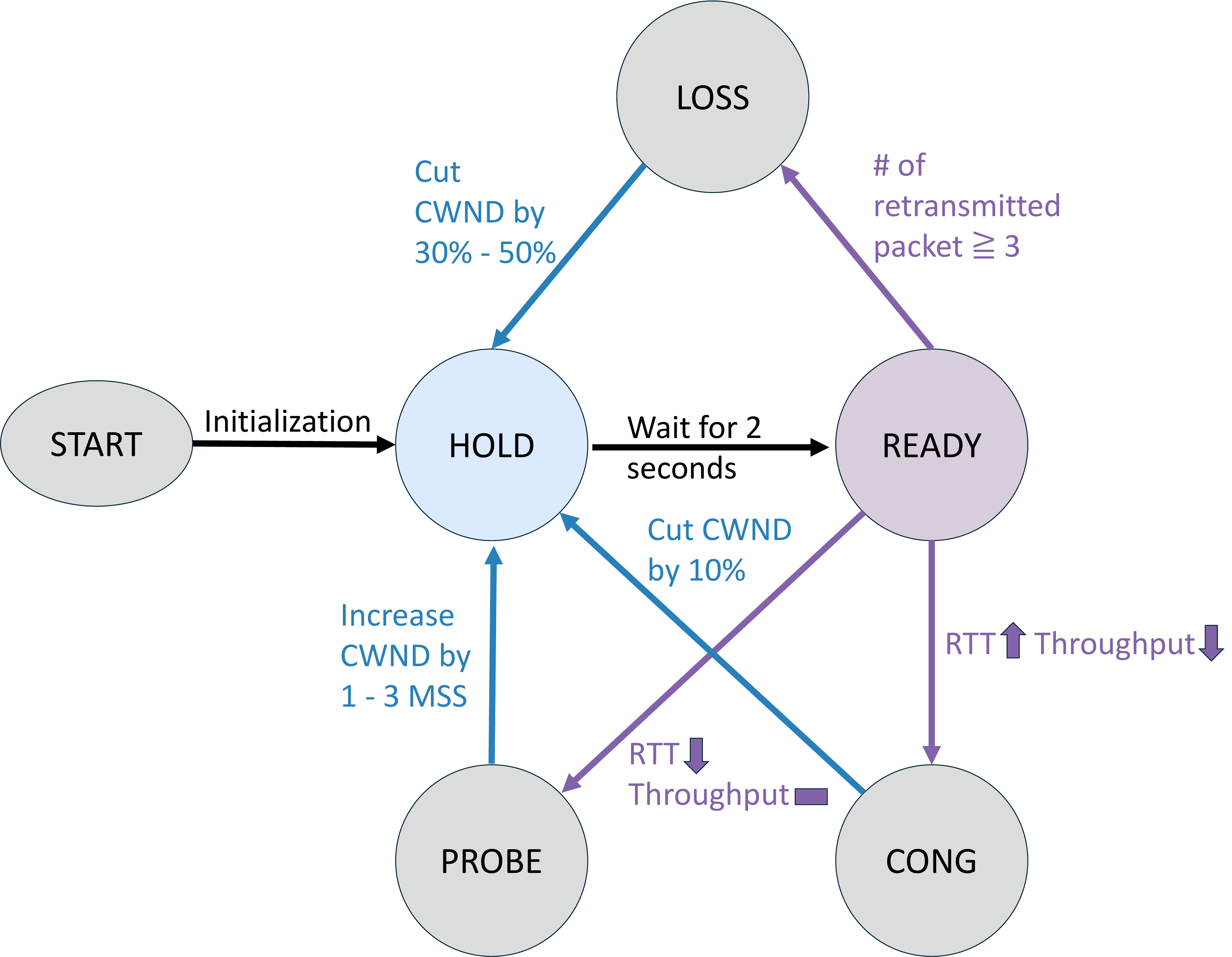}
    \vspace{-6mm}
    \caption{A finite state machine graph of the key decisions and actions of the LLM generated algorithm.}
    \label{fig:fsm}
    \vspace{-3mm}
\end{figure}

\clearpage
\begin{table}[t!]
\centering
\footnotesize
\renewcommand{\arraystretch}{1.25}
\setlength{\tabcolsep}{6pt}
\begin{tabularx}{\textwidth}{>{\bfseries}l|
  >{\columncolor{blue!5}}X|
  >{\columncolor{green!5}}X|
  >{\columncolor{orange!10}}X}
\textbf{Aspect} & \textbf{LLM-Inspired CC} & \textbf{Bbr} & \textbf{Reno} \\ \hline

Congestion Handling &
• Finite-state heuristic \newline
• Adaptive loss cut, conditioned probing, conservative hold \newline
• Finer congestion handling \newline
- Small cwnd reductions around 10\% &
• Model-based \newline
• Estimates bottleneck bw + min RTT &
• Loss-based \newline
• AIMD \\ \hline

Loss Handling &
• Two-step cut: 25\% (mild), 50\% (heavy) \newline
• Reset \texttt{ssthresh} \newline
• 2s cooldown &
• Loss does not directly cut cwnd \newline
• Driven by pacing gain cycle &
• Halve cwnd on loss \newline
• Reset \texttt{ssthresh}, fast recovery \\ \hline

Probing / Growth &
• Add $+\Delta$ MSS only if stable \newline
– No retransmits, stable RTT, $\uparrow$ or steady throughput &
• Periodic gain cycling \newline
– ($+25\%$, $-25\%$) over 8 RTTs &
• +1 MSS per RTT in CA \\ \hline

RTT Sensitivity &
• Explicit RTT gate: $|dRTT|\le\varepsilon$ &
• Maintains min RTT filter \newline
• Probes regardless of queues &
• Ignores RTT growth \newline
• Reacts only to loss \\ \hline

State Machine &
• START \newline
• HOLD, READY \newline 
• LOSS, CONG, PROBE &
• Startup, Drain \newline
• ProbeBW, ProbeRTT &
• SS, CA, FR/FRR, RTO \\ \hline

Timers &
• 2s wait after any action &
• Probe pacing per RTT \newline
• ProbeRTT every $\sim$200ms &
• None  \\ \hline

Delay / Throughput Tradeoff &
• Conservative \newline
• Avoids probing if RTT rising &
• Aggressive probing \newline
• May inflate queues &
• Queues grow until drop \\
\end{tabularx}
\caption{Comparison of LLM-Inspired CC with Bbr and Reno.}
\label{tab:llm_bbr_newreno}
\end{table}
\section{Comparison between LLM generated CC, Bbr and Reno}

In this part, we highlight the differences between the LLM-generated congestion control algorithm and two widely used traditional CCAs, namely Bbr and Reno. To make the comparison clear, we examine seven key aspects that capture their underlying design philosophies and operational behaviors, including principles, loss handling, probing strategy, RTT sensitivity, state machine, timer usage, and delay/throughput trade-offs. The structured comparison in Tab.~\ref{tab:llm_bbr_newreno} illustrates how the LLM-inspired approach departs from conventional loss-based and model-based schemes, emphasizing its gated probing and adaptive loss response.

\end{document}

%% file: 1intro.tex
\vspace{-2mm}
\section{Introduction}
\vspace{-2mm}

TCP congestion control remains one of the most critical yet challenging aspects of the network transport layer design. 
At its core, congestion control is modeled as a goal-directed control loop that, at each sampling instant, the sender is required to choose the next Congestion Window (CWND) that best balances latency and throughput using only sender-visible network information. 

Despite multi-decades research~\cite{Jacob1988,vegas1995,newreno2004,cubic2008,dong2015pcc,cardwell2017bbr}, existing congestion control algorithms (CCAs) all share a fundamental limitation: they rely on hardcoded, static rules for CWND adjustment ~\cite{congestionreview22024,congestionreview12025}. For instance, TCP Reno follows an additive-increase/multiplicative-decrease (AIMD) rule, while CUBIC employs a fixed cubic growth function.
This rigidity leads to sub-optimal performance in modern, highly dynamic settings, such as 5G cellular networks~\cite{prakash2025tcp,xie2020pbe} and satellite links \cite{lai2025leocc,zhao2025satpipe}, where link quality and bandwidth can fluctuate rapidly, making a fixed policy a poor fit.
For instance, these handcrafted CCAs would misinterpret a transient RTT spike from wireless jitter as congestion and reduce CWND unnecessarily, hurting the throughput.

Recently, advances in large language models (LLMs) have demonstrated surprising generalization capabilities \cite{berti2025emergentabilitieslargelanguage, matarazzo2025surveylargelanguagemodels}. As these models scale, they often manifest emergent abilities, such as logical reasoning, planning, and adaptation, that were not even explicitly trained on \cite{valmNEURIPS2023}. Through in-context learning and few-shot prompting, they can rapidly adapt to new tasks and domains without retraining. Moreover, recent work \cite{fu-etal-2025-preact} has demonstrated LLM's planning capacity: integrating contextual signals and constraints to generate multi-step action sequences that achieve goals under uncertainty and across heterogeneous environments.  

Congestion control shares similar characteristics, functioning as a sequential decision-making problem under dynamic network conditions \cite{Jay2019InternetCCRL}. This motivates us to investigate whether an LLM can serve as the decision rulemaker in this control loop, which directly proposes the next CWND to achieve low latency and high throughput in different network conditions. Our hypothesis is that the LLM's capacity for complex, context-aware reasoning might enable it to move beyond static, handcrafted rules and instead dynamically adapt CWND decisions timely and effectively.

Although the high-level idea of using LLM for congestion control is appealing, many open questions remain unanswered.
For instance, what network information should be exposed to the model to enable effective CWND adjustments? To what extent can an LLM interpret such signals to achieve an improved throughput–latency tradeoff? 
Are techniques like in-context learning and prompt tuning, etc., adequate with minor modification? Or is a clean-slate design necessary? Can an LLM-based policy ensure fairness across multiple flows, a long-standing objective in transport layer?

In this paper, we conduct an emulation-guided study to answer these open questions.
We first derisk the design: {\it under a static network with a single flow, can an LLM-based approach achieve latency–throughput tradeoffs comparable to, or even surpassing, those of existing congestion control algorithms?}
We start with granting the LLM limited freedom, allowing it to modify CWND during the congestion avoidance phase while adopting TCP-NewReno for initialization and packet loss handling (\S\ref{sec:baseline}).
This setup allows us to isolate the role of the LLM on a single yet crucial congestion avoidance phase, enabling us to examine three key questions: {\it when} to invoke the LLM (\S\ref{sec:trigger_basline}), {\it what} information to provide (\S\ref{sec:input_design_baseline}), and {\it how} to formulate the LLM instructions (\S\ref{sec:instruct_baseline}).

Building on the insights gained from these explorations, we next 
extend our study to three dynamic network traces to evaluate the algorithm’s generalizability (\S\ref{ss:generalizability}).
Our deep-dive investigation (\S\ref{ss:dee_dive_baseline}) on the delay, CWND, and queue size over time demonstrate this vanilla LLM-based policy, namely, TCP-LLM-L (where L stands for limited), can self-adaptively balance the RTT and throughput and achieve better performance than 11 baselines (including Bbr, Cubic, etc.), particularly on highly dynamic network conditions. In the meantime, we observe that the limited freedom granted to the LLM leads it to favor decreases or no-change actions on CWND control, rarely attempting increases, which
caps its ability to fully exploit available bandwidth.

This observation motivates us to generalize LLM-based control across multiple congestion control phases (\S\ref{sec:generalized}).
To achieve this goal, we propose to use ACK count as another crucial input to LLM (\S\ref{sec:trigger_generalized}) and design an iterative prompt refinement policy to gradually optimize the LLM's prompt (\S\ref{sec:input_design_generalized}).
We examine the performance of this new congestion control policy, namely, TCP-LLM-G (where G stands for generalized) in three real-world, dynamic network traces.
We further unpack this TCP-LLM-G's CWND control operations and compare against the baseline TCP-LLM-L (\S\ref{ss:deep_dive_generalized}). The results show that the generalized action
space together with the ACK-based trigger makes TCP-LLM-G more adaptable across diverse topologies, consistently balancing latency and throughput better than TCP-LLM-L.

Moving beyond the one-to-one topology, we further explore multi-sender competing for a single link, with the goal of examining the fairness among these competitors in both homogeneous and hybrid network setups over all four real-world network traces (\S\ref{s:fairness}).
Moreover, we use GPT-5 ~\cite{openai2025gpt5systemcard} to translate these LLM-guided congestion control policies into formalized algorithms (\S\ref{s:interpretability}), with the goal of figuring out their difference with existing hard-coded CCAs, and further better understanding why they achieve better performance in low-latency and high-throughput.

Through this step-by-step, emulation-guided exploration, we systematically investigate the feasibility and challenges of leveraging large language models for network congestion control. Our findings not only reveal the potential of LLM in congestion control but also establish foundational principles for future research. Specifically, our key contributions and core takeaways from this exploration are as follows:

\begin{itemize}
\item {\it LLM reasoning in CC}: LLMs can effectively reason and make decisions (even with zero-shot prompting) when provided with relevant historical context and instructions consistent with their pre-training.
\item {\it Degree of control freedom}: Maximizing an LLM's generalization in congestion control requires granting it ample freedom in its decision-making space.
\item {\it Triggering mechanism}: An efficient triggering module is crucial. Latency-based triggers were too conservative, while ACK-based triggers offer better adaptability but depend heavily on threshold tuning.
\item {\it Fairness}: LLM-based congestion control achieves fairness among LLM flows, but is outcompeted by the aggressive CWND growth of conventional CCAs.
\end{itemize}

%% file: 2related.tex
\vspace{-4mm}
\section{Related Work}
\vspace{-2mm}
We review related works in this section.

\begin{figure*}[t]
  \centering
  \includegraphics[width=\textwidth]{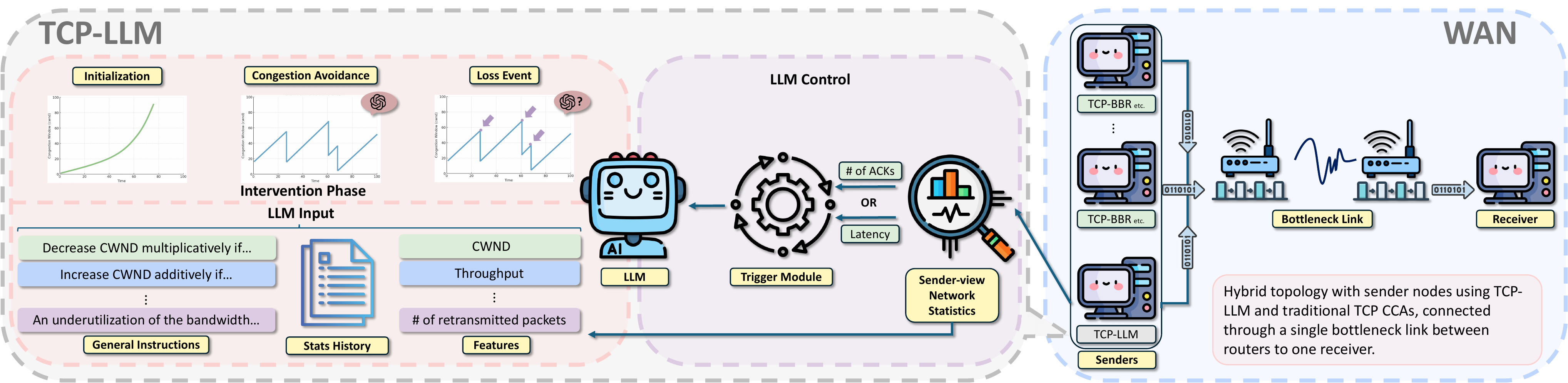}
  \caption{Overview of the design of \textbf{TCP-LLM} in a hybrid WAN topology, which is developed following main questions: (1) \textbf{How much freedom should LLM be granted?} (Sec~\ref{sec:baseline} VS. Sec~\ref{sec:generalized}) Probed via abstraction of the congestion control problem into three phases and deploying LLM in different phases. (2a) \textbf{When to invoke the LLM?} (Sec~\ref{sec:trigger_basline}, Sec~\ref{sec:trigger_generalized}) Addressed through building latency-based and ACK-based trigger module. (2b) \textbf{What information does LLM need?} (Sec~\ref{sec:input_design_baseline}, Sec~\ref{sec:input_design_generalized}) Explored by comprising three different components of general instructions, network statistics features, and statistics history. }
  \label{fig:pipeline}
  \vspace{-3mm}
\end{figure*}

\vspace{-3mm}
\subsection{Learning-Based Congestion Control}
\vspace{-1mm}

Early works on machine learning for congestion control focused on automating the design of TCP algorithms to break free from fixed set of heuristic rules. A landmark example is Remy~\cite{Remy}, which uses an offline model-based optimizer to synthesize a congestion control algorithm designed to network assumptions and performance objectives. PCC\cite{dong2015pcc} and PCC Vivace\cite{dong2018pcc-vivace} take a further step forward towards online learning-driven congestion control that can be adjusted in real time through utility-guided experiments. Their experiments demonstrated the potential of enabling real time adjustments in CCA through learning-based approaches. 

With advances in deep reinforcement learning (DRL), more recent works apply DRL to congestion control, viewing the CCA as an agent that observes network state and adjusts the sending rate or window for actions. Pioneer work Aurora~\cite{Jay2019InternetCCRL} employs PPO (Proximal Policy Optimization) to train a neural network for sending rate control. Later on, several works~\cite{fang2019r3net, sivakumar2019MVFSTRL, xu2019drl-cc, emara2020eagle, orca2020} exploit different DRL techniques. These approaches achieve desirable throughput and latency under the conditions they are trained on, but because they are typically trained offline in simulated networks, they often fail to generalize to unseen environments, as pointed out in Mutant~\cite{pappone2025mutant}. 
Mutant instead ensembles multiple existing congestion control algorithms. It continuously monitors their performance under current conditions and dynamically selects CCA rules, effectively “mutating” its behavior in real time. 
Building on Mutant’s philosophy of online adaptability, we investigate how LLMs’ generalization capacity can be used to create a CCA agent that dynamically balances throughput and latency under varied network scenarios without the restrictions of existing predefined CCA rules.

\vspace{-3mm}
\subsection{Applying LLM to Networking Domain}
\vspace{-1mm}
The problem-solving capability of LLMs has motivated their application to networking problems. NetLLM~\cite{wu2024netllm} was proposed as a framework to investigate how a single pre-trained LLM can be adapted to understand and handle diverse networking tasks, such as video streaming rate control. It's SOTA performance across multiple tasks suggests that pretraining has embedded meaningful networking knowledge within these models. We view the problem of applying LLMs to new domains as a challenge of asking the right question in the right way. Motivated by this perspective, we conduct an exploratory study to investigate the requirements and design considerations for employing an LLM for congestion control that can break out the boxes of existing CCAs.

We are not the first applying LLM to congestion control.
There are several works \cite{shrestha2024tcp-llm, he2025llm-generated-algo} have already took the first leap, but they have primarily focused either on allowing the LLM to select an existing CCA~\cite{shrestha2024tcp-llm} or to augment the rules of a specific CCA~\cite{he2025llm-generated-algo}. We instead take a ground-up approach to study whether an LLM can serve directly as the controller and rule-maker, generating values to adjust sending rates instead of merely selecting or modifying an existing algorithm. We envision our exploration results and takeaways can spawn many new ideas on leveraging LLM for congestion control.

%% file: 3overview.tex
\vspace{-3mm}
\section{Defining Protocol States and LLM Intervention in Congestion Control}
\label{s:rethinking}
\vspace{-2mm}
To precisely bound where the LLM may intervene, we follow existing CCAs and model congestion control as a discrete set of protocol states: {\it initialization}, {\it congestion avoidance}, and {\it packet loss}, with the latter defined by 3-duplicate ACKs or timeout signals~\cite{newreno2004}. \footnote{Note that in most CCAs, there is a fourth phase: fast retransmit/fast recovery. In our study, we abstract this into a single packet loss phase, which covers both loss handling and recovery.} Each state carries a distinct control objective and risk profile for adjusting CWND. 

A central methodological question then arises: {\it how much freedom should the LLM be granted in this multi-state control loop}? Blindly granting broad control across all protocol states risks instability and even violations of TCP invariants. 
These risks arise under partial observability (sender-only view), delayed and asynchronous feedback, and non-stationary cross-traffic that would make RTT/throughput signals drift and become ambiguous. 
For instance, by the time the LLM reacts to an observed loss, the congestion window it decided to shrink may no longer reflect the current state of the network, causing oscillations.
Conversely, if the model is over-constrained, potential gains from LLM's generalization ability may be forfeited and the problem might be reduced to a trivial policy. To further understand this trade-off, we organize our study around two guiding questions that together determine the LLM’s reasonable scope of intervention: 

\begin{itemize}
    \item {\it Q-(a): When should the LLM be triggered?}
    \item {\it Q-(b): What information does the LLM require to reason about CWND adjustment effectively?}
\end{itemize}

For question (a), triggering LLMs should balance the congestion control performance with practicality and cost. Triggering LLM too frequently is usually not practical: it incurs API call cost, and may cause asynchronous actions from observations due to model inference latency. What we need, therefore, are reliable signals that indicate a meaningful opportunity for intervention. In other words, the challenge becomes finding well-defined moments in the congestion control process where sender-visible feedback is strong enough to justify invoking the LLM model.
We present the design in response to this challenge in \S\ref{sec:trigger_basline} and \S\ref{sec:trigger_generalized}, respectively.

For question (b), we proceed by specifying what the LLM model can reliably observe at the sender in real-world deployments and how that information should be represented. We initially consider five candidate features: {\it CWND}, {\it RTT}, {\it throughput}, {\it number of retransmitted packets}, and {\it network topology}. We exclude topology because router-level conditions along the path cannot be directly observed or reliably maintained at the sender side. This yields a compact, sender-only feature set: {\it CWND as the primary control variable; RTT and its short-term trend as proxies for latency; throughput and its trend as indicators of bandwidth utilization; and the recent count of retransmissions as a direct signal of loss.} 
We discuss the specific use of these features in \S\ref{sec:input_design_baseline} and \S\ref{sec:input_design_generalized}.


Based on the above analysis, we plan a staged approach to derisk the use of LLM in congestion control.
We first constrain the LLM’s intervention to the congestion avoidance state (\S\ref{sec:baseline}), where adjustments are incremental and the risks of destabilizing the flow are relatively contained.
This allows us to evaluate whether the model can meaningfully achieve on-par or improvement over static policies in balancing latency and throughput without jeopardizing correctness.
We then extend its scope to other states (\S\ref{sec:generalized}). This incremental roadmap allows us to assess trade-offs systematically and establish guardrails before granting the LLM broader control.

%% file: 4limited.tex
\vspace{-3mm}
\section{De-risking with Boundaries: LLM in Congestion Avoidance}
\label{sec:baseline}
\vspace{-2mm}

\begin{figure}[t]
  \centering
  \includegraphics[width=\columnwidth]{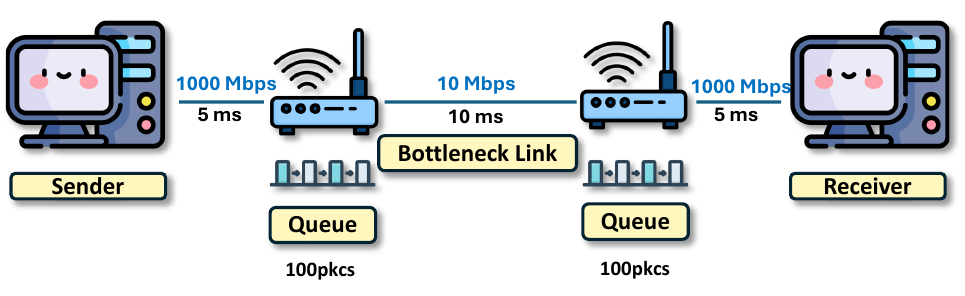}
  \vspace{-8mm}
  \caption{The topology used in LLM exploration.}
  \vspace{-5mm}
  \label{fig:one2one_topo}
\end{figure}

We begin by granting the LLM limited freedom, allowing it to modify CWND only during the congestion avoidance phase. We start with the congestion avoidance phase because it is not only the longest and most frequently encountered phase in a congestion control run, but also the phase that offers the greatest flexibility for LLM intervention. 

To explore the LLM-based control for congestion avoidance, we implement a two-node network topology inter-connected with two routes using NS-3\footnote{https://www.nsnam.org/releases/ns-3-43/}, as shown in Fig.~\ref{fig:one2one_topo}.
This controlled setup enables us to isolate and directly observe the effects of our design choices.
For the initialization and loss phases, we directly employ the mechanisms of TCP-NewReno. In addition, during the congestion avoidance phase, if LLM is decided not to be triggered upon receiving an ACK, we also employ the mechanisms of TCP-NewReno. 

We choose to build upon TCP-NewReno because it is both widely adopted and analytically simple. Its AIMD rules produce straightforward, easily recognizable patterns in traces, which allows us to isolate the role of the LLM without interference from more complex behaviors such as Bbr’s probing. 
To prevent the system from reverting to the initialization phase when the LLM generates a new CWND, we enforce the condition that the resulting SSTHRESH is equal to the updated CWND. 
We use GPT-4o-mini \cite{openai2024gpt4ocard} as the LLM.


\vspace{-3mm}
\subsection{When to Invoke the LLM?}
\label{sec:trigger_basline}
\vspace{-1mm}
An essential design question is determining when to invoke the LLM to adjust CWND. This requires a trigger module that continuously monitors network conditions and decides whether an adjustment should be applied. In conventional loss-based congestion control algorithms such as TCP-NewReno, congestion is detected through packet loss signals. However, these signals are inherently delayed, as the receipt of 3-duplicate ACKs indicates that congestion has already occurred. To enable earlier intervention, we adopt a latency-based approach inspired by existing delay-sensitive congestion control schemes (e.g., Bbr \cite{cardwell2017bbr}), which treat RTT growth as an early indicator of congestion rather than waiting for loss. Specifically, we monitor the latency of the most recently sent segment, and if it exceeds a predefined threshold, we trigger the LLM to adjust CWND.

Choosing an appropriate threshold for congestion detection is inherently a critical challenge. Rather than relying on arbitrary, hand-tuned values that may not generalize across different network conditions, we adopt a more adaptive approach. We first run a short simulation using standard TCP-NewReno and record the latency at the moment when the first packet loss occurs (for example, in the one-to-one static topology it is 160ms), using it as an estimate of the latency level associated with true congestion. One thing to note is that this estimation can be rerun if the network condition is detected drastically changed. To set the final triggering threshold, we scale this baseline latency by a factor $\alpha$ (where $\alpha < 1$), allowing the system to detect congestion proactively before losses occur. We conduct an ablation study on the choice of $\alpha$ and its influence on performance in \S\ref{sec:latency_trigger_ablation}. In essence, the choice of $\alpha$ serves merely as a tradeoff parameter between cost and performance, rather than a requirement for the system’s correctness. In our experiment, we chose $\alpha = 70\%$ for better performance with limited cost.

To ensure that the impact of an LLM-triggered adjustment is accurately reflected in subsequent network statistics, we introduce an additional timing constraint. Since changes to CWND require time to propagate through the network and influence measured metrics, invoking the LLM immediately after an adjustment risks basing decisions on outdated or transient observations. To mitigate this effect, we enforce a waiting period of two seconds between consecutive LLM activations. Consequently, the LLM is triggered only when the observed latency exceeds a threshold and at least two seconds have elapsed since the most recent intervention.

\vspace{-3mm}
\subsection{What to Tell the LLM?}
\label{sec:input_design_baseline}
\vspace{-1mm}

To leverage the pretrained knowledge of the LLM for congestion control, we should determine what information should be provided to enable meaningful adjustments to CWND. Following the standard structure of LLM prompting, the input consists of two components: {\it general instructions}, fixed across all queries, and {\it input features}, which vary with current network conditions. This separation ensures consistent guidance while enabling responsiveness to network dynamics.

We first specify generalized instructions that define congestion and set the dual objectives of maximizing throughput and minimizing latency. These instructions remain fixed during feature selection, where we identify the most informative network statistics. Once the relevant features are chosen, the instructions are refined to align with them, ensuring the LLM’s guidance is both targeted and effective.

Specifically, we conduct stepwise feature selection by providing the LLM with the latest statistics of the four features: CWND, RTT, throughput, number of retransmitted packets. We find that the number of retransmitted packets does not appear to be a critical feature in this setup because the latency-based trigger module, which operates preemptively to mitigate congestion and to reduce the likelihood of packet loss.

After securing the final three features, we turn to refining the general instructions to further improve throughput and reduce latency. We experiment with alternative ways of formulating the rules for CWND adjustment under different network conditions, as shown in Tab.~\ref{tab:math_prompt_scheme} and Tab.~\ref{tab:natural_prompt_scheme}, which illustrates both formulations: 
\begin{itemize}
    \item \textbf{Math}, where the changes are expressed using mathematical formulas following AIMD.
    \item \textbf{Natural}, where they are described in natural language without specifying the exact amount of adjustment. 
\end{itemize}

\begin{table}[t]{
\begin{tcolorbox}[
    colframe=blue!50!black,
    colback=white,
    coltitle=white,
    fonttitle=\bfseries,
    title=Math Prompt Scheme,
    boxrule=1.2pt
    ]

{\footnotesize
\vspace{-2mm}
\textbf{System Prompt:}\\
You are an expert congestion controller in wide area networks. 
Inputs: \{Inputs Math Symbols: last\_cwnd := c\_\{t-1\} ...\}. Update rules (produce next\_CWND = c\_\{t+1\}, next\_SSThreshold = s\_\{t+1\}):\\
(1) Congestion: \\
If $r_t > r_{t-1}$ and $T_t \not> T_{t-1}$, 
choose a multiplicative decrease factor $\beta \in (0.5, 1)$, 
and set $c_{t+1} = \max\!\left(\lfloor \beta \cdot c_t \rfloor, MSS \right)$...
}
\vspace{-2mm}
\end{tcolorbox}
}
\vspace{-4mm}
\caption{A sketch of math prompt.}
\label{tab:math_prompt_scheme}
\end{table}

\begin{table}[t]{
\begin{tcolorbox}[
    colframe=blue!50!black,
    colback=white,
    coltitle=white,
    fonttitle=\bfseries,
    title=Natural Prompt Scheme,
    boxrule=1.2pt
    ]

{\footnotesize
\vspace{-2mm}
\textbf{System Prompt:}\\
You are an expert congestion controller in wide area networks.
Based on the following input parameters: \{Input Explanations: last\_cwnd: The previous congestion window size...\}. Modify cwnd and ssthreshold according to the following rules:\\
1. Decrease cwnd multiplicatively if there might be a congestion.\\
2. Increase cwnd additively if there might be underutilization of the bandwidth...
}
\vspace{-2mm}
\end{tcolorbox}
}
\vspace{-4mm}
\caption{A sketch of natural prompt.}
\label{tab:natural_prompt_scheme}
\vspace{-3mm}
\end{table}

\begin{table}[t]
\centering
\small
\begin{tabular}{l c c c c c}
\hline
\multicolumn{2}{c}{} & \multicolumn{2}{c}{RTT (ms)} & \multicolumn{2}{c}{Throughput (Mbps)} \\
\cline{3-6}
Type & Few-shot? & Avg. & Std. & Avg. & Std. \\
\hline
\multicolumn{6}{l}{\textit{(a) Instruction Scheme / Few-shot}} \\
Math     & - & 125.94 & 1.15 & 9.99 & 0.00 \\
Natural  & - & \textbf{92.83} & 3.54 & \textbf{9.99} & 0.00 \\
Natural  & + & 113.21 & 2.02 & 10.00 & 0.00 \\
\hline
\multicolumn{6}{l}{\textit{(b) History Length $H$} experiment} \\
$H=1$ &  & 110.57 & 5.78 & 9.99 & 0.00 \\
$H=2$ &  & 95.99 & 4.49 & 9.99 & 0.00 \\
$H=3$ &  & 95.30 & 3.18 & 9.99 & 0.00 \\
$H=4$ &  & \textbf{92.83} & 3.54 & 9.99 & 0.00 \\
\hline
\end{tabular}
\caption{Performance results: (a) Instruction scheme and zero-shot vs. few-shot, and (b) History length $H$ experiment. Notice that because the LLM was run with zero temperature (deterministic outputs), under a restricted action space, and in a stable-bandwidth network environment, the variability across repeated runs was negligible. Consequently, the reported standard deviations round to 0.00 when shown to two decimal places.}
\vspace{-5mm}
\label{tab:combined_ablation}
\end{table}

The guiding rules direct the LLM to first assess whether congestion is present. If congestion is detected, the LLM is instructed to reduce CWND; otherwise, it may either increase CWND when bandwidth appears underutilized or leave it unchanged when performance is satisfactory. The performance comparison of these prompt designs is presented in part (a) in Tab.~\ref{tab:combined_ablation}. We observe that the \textbf{Math} scheme leads the LLM to closely emulate TCP-NewReno’s behavior, strictly adjusting CWND according to the formulas, and thereby resulting in performance similar to TCP-NewReno. This indicates that the intervention does not leverage the LLM’s generalization ability, and is therefore not meaningful, given that triggering the LLM incurs higher cost than directly using TCP-NewReno.

In contrast, the \textbf{Natural} scheme yields improved performance, which we attribute to its closer alignment with the LLM’s pretraining data. Since the word-based instructions avoid explicit numerical constraints, they leave greater flexibility for the LLM to adjust CWND more effectively. During this process, we observe that the LLM occasionally outputs adjustments of less than one segment, which are not meaningful; therefore, we add explicit rules instructing the model to generate only valid values that comply with TCP-safe guardrails. We include the final prompts in Appendix Tab.~\ref{tab:gpt_prompts_baseline}.

Furthermore, to provide sufficient context for detecting trends in network dynamics, we extend the input with multiple rounds of historical statistics. For example, a history length of 3 includes the statistics from the 3 most recent segments. From part (b) in Tab.~\ref{tab:combined_ablation}, we observe that as we increase the length of the history, the average latency of the system reduces, which is indicating that a longer history might help the LLM gain a better understanding of the network condition.


\begin{figure*}[t]
  \centering
  \includegraphics[width=\textwidth]{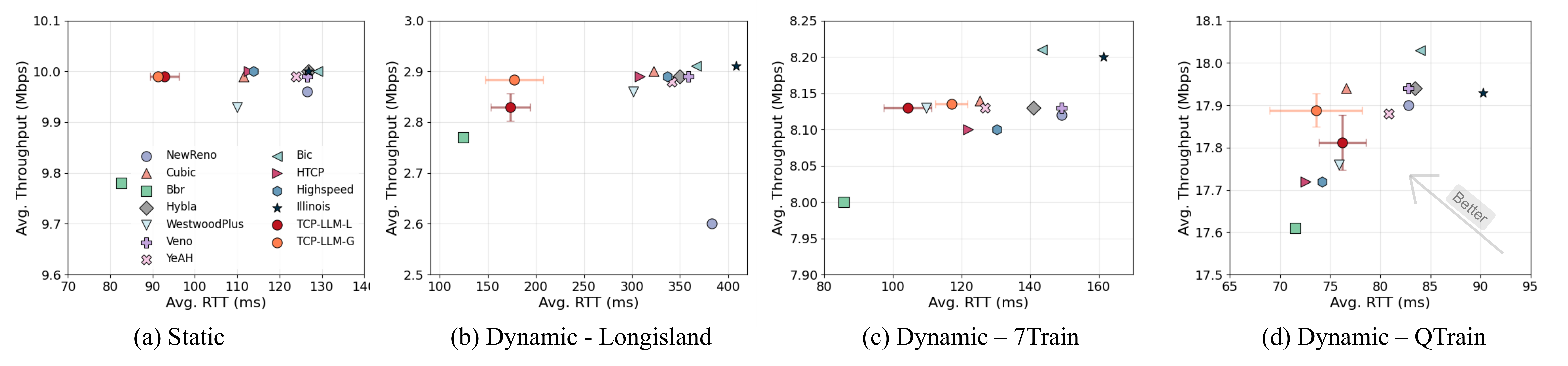}
  \vspace{-6mm}
  \caption{Performance comparison of \textbf{classic CCAs} and \textbf{TCP-LLM limited design} (TCP-LLM-L, \S\ref{sec:baseline}) and \textbf{TCP-LLM general design} (TCP-LLM-G, \S\ref{sec:generalized}) on four different links: \textbf{Static}, \textbf{Dynamic-Longisland}, \textbf{Dynamic-7Train}, and \textbf{Dynamic-QTrain}.}
  \label{fig:baseline_limited}
  \vspace{-5mm}
\end{figure*}

\vspace{-3mm}
\subsection{How to Instruct the LLM?}
\label{sec:instruct_baseline}
\vspace{-1mm}
\textbf{Zero-Shot or Few-Shot}? We next investigate whether the in-context learning ability of the LLM can be leveraged to improve performance by incorporating few-shot examples into the prompt. To obtain reference examples, we ran the TCP-NewReno simulation pipeline, and because we were not able to identify the gold few-shot example, we randomly chose one instance of CWND adjustments. A performance comparison between the zero-shot and few-shot schemes is presented in Part (a) in  Tab.~\ref{tab:combined_ablation}. Contrary to our expectation, the inclusion of few-shot example results in relatively degraded performance. Closer inspection reveals that, under the few-shot scheme, the LLM tends to directly reproduce values similar to the provided example rather than inferring new adjustments based on the current input. By contrast, the zero-shot setup allows the LLM greater flexibility to adapt its outputs to varying conditions. We therefore adopt the zero-shot setup for the remainder of our probing experiments.

\vspace{-3mm}
\subsection{Does the System Generalize Across Diverse Network Conditions?}
\label{ss:generalizability}
\vspace{-1mm}
\textbf{Emulation Setups}. We extend our experimental setup to dynamic links, emulating realistic network bandwidth fluctuations to check whether LLM can generalize. 
We achieve this by using three distinct traces, namely, Longisland, QTrain, and 7Train, from the NYU Metropolitan Mobile Bandwidth Trace dataset \cite{mei2020realtime} to our one-to-one topology. 
  
\begin{itemize}
    \item \textbf{Longisland}, characterized by a single spike of high bandwidth amid predominantly low bandwidth.
    \item \textbf{QTrain}, featuring frequent and abrupt fluctuations.
    \item \textbf{7Train}, showing an overall increasing trend with relatively small variations.
\end{itemize}

Each trace simulates bandwidth variations every five seconds over 120-second dynamic network simulations. 
We particularly focus on wireless traces as they inherently exhibit a high degree of variability and unpredictability, posing a significant challenge for congestion control. 

We compared the LLM-based congestion control policy, TCP-LLM-L, which operates only during congestion avoidance, against a broad set of representative CCAs. We include (i) loss-based probing, e.g., NewReno~\cite{newreno2004}, Cubic~\cite{cubic2008}, where congestion signals are inferred primarily from packet drops; (ii) Delay-sensitive control, e.g., YeAH ~\cite{Baiocchi2007YeAH}, Illinois ~\cite{illinois2008}, which emphasizes RTT trends as early indicators of congestion while still accounting for loss; and (iii) model-based pacing, e.g., Bbr~\cite{cardwell2017bbr}, which departs from traditional AIMD heuristics by explicitly estimating bottleneck bandwidth and delay. In this experiment, we set the trigger threshold to 112ms, which is $\alpha = 70\%$ for an estimation of 160ms ran in the static topology. We set the history length to 4.

\begin{figure*}[t]
  \centering
  \includegraphics[width=\textwidth]{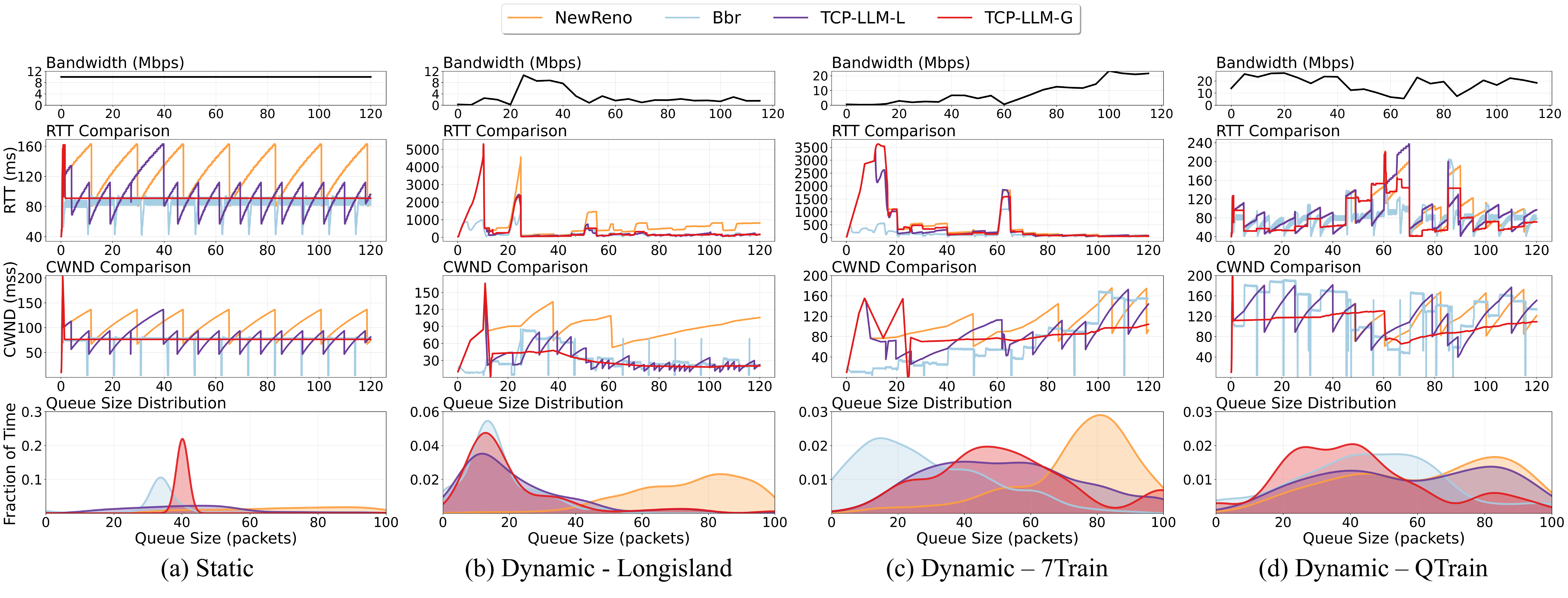}
  \vspace{-9mm}
  \caption{Snapshots of \textbf{\textcolor[HTML]{6A3D9A}{TCP-LLM-L} (\S\ref{sec:baseline})}, \textbf{\textcolor[HTML]{E31A1C}{TCP-LLM-G} (\S\ref{sec:generalized})}, \textbf{\textcolor[HTML]{FEA042}{NewReno}}, and \textbf{\textcolor[HTML]{A6CEE3}{Bbr}}: 
\textbf{Row 1}: Bandwidth changes of the bottleneck link over time.  
\textbf{Row 2}: RTT comparison over time.  
\textbf{Row 3}: CWND comparison over time.  
 \textbf{Row 4}: Bottleneck link router queue size distribution for all 120s.   
}
\vspace{-5mm}
  \label{fig:snapshot_final}
\end{figure*}

\noindent\textbf{Results analysis}. As shown in Fig.~\ref{fig:baseline_limited}, across both static and dynamic network conditions, TCP-LLM-L generally achieves the most balanced trade-off between throughput and latency. While Bbr attains notably lower RTTs, it does so at the expense of substantially reduced throughput compared to other CCAs. In contrast, algorithms such as Cubic and WestwoodPlus sustain relatively higher throughput but incur significantly higher latency. In comparison, TCP-LLM-L delivers near-maximum throughput while maintaining latency that is markedly lower than most loss-based schemes. 

For example, in the static trace shown in Fig.~\ref{fig:baseline_limited}(a), TCP-LLM-L achieves an average throughput of 9.99 Mbps on a 10 Mbps link, similar to loss-based CCAs such as NewReno and Cubic, despite not employing their more aggressive window growth strategies. Yet, it does so with remarkably lower latency, between 17\% and 30\% less than these standard CCAs, while incurring only a modest 9\% increase compared to Bbr, which is known for its latency-oriented design. The advantage becomes even more pronounced in the Longisland trace, TCP-LLM-L reduces latency by approximately 45\%–70\% relative to most rule-based CCAs, while sustaining throughput that is nearly indistinguishable from theirs.
This balance demonstrates that TCP-LLM-L avoids the extremes of underutilization and excessive queueing, yielding competitive performance across diverse links.

For traces such as 7Train and QTrain shown in Fig.~\ref{fig:baseline_limited}(c) and Fig.~\ref{fig:baseline_limited}(d) respectively, which exhibit gradual increases or frequent and abrupt fluctuations, TCP-LLM-L demonstrates less pronounced advantages compared to the Static and LongIsland cases. This is because window growth in TCP-LLM-L relies primarily on the underlying TCP-NewReno process; the LLM itself contributes little to proactive bandwidth probing, and its latency-triggered design emphasizes delay reduction rather than fully exploiting available capacity. For instance, in the 7Train scenario, TCP-LLM-L reduces latency by at least 5\% compared to most rule-based CCAs, while incurring only a marginal throughput loss of about 1.2\%. However, it incurs roughly 20\% higher latency than Bbr, while still delivering about 2\% higher throughput, increasing from 8.00 Mbps to 8.13 Mbps on a link with an average capacity of around 8.20 Mbps. As a result, TCP-LLM-L still performs reasonably well in these dynamic scenarios, though its benefits are not as prominent as in the Static and LongIsland traces.

\begin{figure*}[t]
  \centering
  \includegraphics[width=\textwidth]{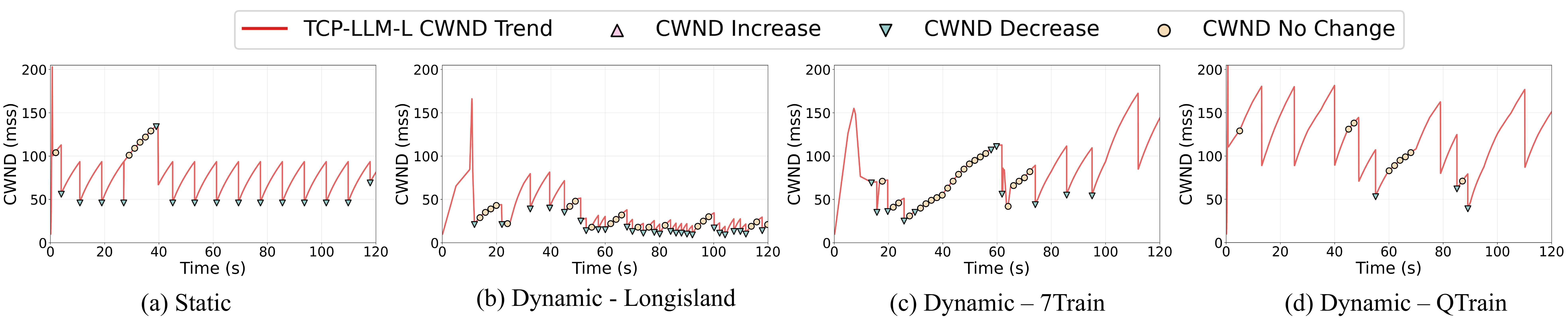}
  \vspace{-8mm}
  \caption{LLM behavior in \textbf{TCP-LLM-L(\S\ref{sec:baseline})} over 4 topologies: (a) \textbf{Static}, (b) \textbf{Dynamic-Longisland}, (c) \textbf{Dynamic-7Train}, and (d) \textbf{Dynamic-QTrain}.}
  \label{fig:llm_behav_l}
  \vspace{-3mm}
\end{figure*}

\vspace{-4mm}
\subsection{Deep-Dive Investigation: Unpacking LLM Congestion Control Strategies}
\vspace{-1mm}
\label{ss:dee_dive_baseline}

To gain deeper insight into how the LLM behaves under varying conditions, we plot LLM's control over time. The goal is to examine whether the model can promptly adapt its decisions in response to abrupt bandwidth shifts at a closer scale, besides comparing the average performance metrics. 

\noindent\textbf{Static.} As shown in Fig.~\ref{fig:snapshot_final}(a), we begin with the static network condition where the link bandwidth is consistent over time. We observe that Bbr achieves significantly lower RTT, below 89ms for most of the time, yet sacrificing throughput to the large extend as exhibited in Fig.~\ref{fig:baseline_limited}(a), whereas TCP-NewReno maintain relatively high CWND values to saturate bandwidth, which in turn leads to higher average latency of 126ms. TCP-LLM-L maintains a CWND similar to Bbr by consistently reducing CWND when needed, thereby achieving both lower latency and higher throughput.

\noindent\textbf{Longisland.} Next, we analyze the reason behind TCP-LLM-L's advantage in average latency and throughput in the Longisland trace in Fig.~\ref{fig:snapshot_final}(b), which has one significant spike from 2Mbps to 10Mbps around 20–50 seconds. Starting from 20 seconds, NewReno rapidly adapts to the increase in bandwidth. But the pre-increase CWND of 90MSS was too high for a low bandwidth of 2Mbps, so it sends too much packets before the bandwidth rises, incurring large increases in RTT to around 4500ms at 25 seconds. As the spike disappears at 50 seconds, NewReno does not adapt to the lower bandwidth and continue assigning high CWND value, causing the highest latency of 1420ms. However, we observe that Bbr and TCP-LLM-L are more conservative in their CWND increases and are able to adapt to both the abrupt increase and the abrupt decrease timely, achieving a lower latency of around 100ms starting from 50 seconds onward. In addition, we note that Bbr attempts to probe the network periodically after 50 seconds, then drops CWND to only 4 MSS after failed probes, causing it to lose throughput compared to TCP-LLM-L, which safely operates with lower CWND values around 20MSS. From the queue size distribution plot in row 4, we can see that Bbr and TCP-LLM-L clearly has a higher fraction of low queue size than NewReno, which further demonstrate their latency advantages.

\noindent\textbf{7Train.} We then turn to the 7Train trace in Fig.~\ref{fig:snapshot_final}(c), in which TCP-LLM-L shows the most evident advantage of latency in the first 40 seconds. We observe in the first 40 seconds, when the network bandwidth is very limited at 1 or 2Mbps, all CCAs except for Bbr attempt to lower CWND to adapt to the bandwidth from their high initialization of 150MSS, whereas Bbr starts with a low CWND. TCP-LLM-L decreases CWND while NewReno still increases throughout these 40 seconds, thus TCP-LLM-L achieves lower latency than NewReno. Bbr again sacrifices throughput for lower latency. This is shown in the queue size distribution plot in row 4, the bottleneck queue for Bbr is empty of 1\% of the timestamps, indicating an underutilization of bandwidth. When the bandwidth gradually increases starting at 60 seconds, all CCAs adapt well to this slow change.

\noindent\textbf{QTrain.} Finally, we evaluate each CCA's performance under the highly abrupt bandwidth fluctuations characteristic of the QTrain trace, as depicted in Fig.~\ref{fig:snapshot_final}(d). Unfortunately, in this particular scenario, TCP-LLM-L does not exhibit the same performance advantage observed in the other three traces. Most CCAs, excluding Bbr, demonstrate similar performance trends, with none showing a direct adherence to the rapid bandwidth fluctuations. Our analysis of snapshots across the four topologies reveals that, in QTrain, the LLM is not triggered as frequently, leading TCP-LLM-L to rely more heavily on its default NewReno behavior and consequently diminishing its expected advantages.

This observation motivates us to revisit the latency-based triggering module. We notice that the bandwidth in the initial segment of the QTrain trace (0 to 45 seconds) is approximately twice the 10 Mbps bandwidth characteristic of our static trace. This significant divergence in available bandwidth lead to an underestimation of network latency by the static-case derived model, and thus preventing the TCP-LLM-L from reaching its pre-determined latency threshold and thereby delaying its intervention. 
This lack of action by the LLM explains the vanishing advantage.

\vspace{-3mm}
\subsection{Conclusion Remarks}
\label{ss:concl}
\vspace{-1mm}


The experiment results are promising so far -- they demonstrate LLM can self-adaptively balance the RTT and throughput, particularly on highly dynamic network conditions. However, we also observe a behavioral bias: despite being provided with general instructions for {\it increase}, {\it decrease}, and {\it unchange}, the LLM consistently opted for decrease or unchange actions across all four topologies, as illustrated in Fig.~\ref{fig:llm_behav_l}. 
We attribute this limited range of actions (decrease or unchange) to the constrained operational freedom granted to the LLM: since it is only triggered when latency is high during the congestion avoidance, the presented network statistics almost invariably indicate congestion. Consequently, the LLM predominantly chooses to reduce or retain the CWND, even though the prompt explicitly allows for increases.

Furthermore, the LLM's decrease actions are stricktly uniform, consistently reducing the congestion window (CWND) by exactly 50\%. We attribute this consistent reduction, a percentage not defined in the explicit instructions, to an inherited bias from its pretraining on classical congestion control algorithms or networking textbooks.

These observations motivate us to grant LLM greater autonomy, both in its engagement across various congestion control phases and within the triggering module (\S\ref{sec:generalized}).



{\it \noindent\textbf{Takeaway}. Restricting LLM intervention to congestion avoidance provides a safe entry point, reducing latency while preserving throughput. Moreover, natural-language prompts outperform math-based ones, aligning with LLM pretraining. Latency-based triggers and history length shape the trade-off: lower thresholds cut latency at some cost to throughput, while longer history improves stability. However, limited freedom makes the LLM favor decreases or no-change, rarely increasing, which caps bandwidth use.}




%% file: 5free.tex
\vspace{-4mm}
\section{Generalized Design: LLM Across Multiple Congestion Control Phases}
\label{sec:generalized}
\vspace{-3mm}

In this setup, we allow the LLM to adjust CWND during both congestion avoidance and upon 3-duplicate ACKs events, giving it more latitude to react to mild loss. Note that in this design, we do not include a separate fast retransmit/fast recovery phase; instead, we let the LLM directly adjust the CWND in response to loss events. Our core hypothesis is that this enhanced freedom will allow the LLM to better leverage its generalization ability to quickly restore stable throughput and effectively mitigate starvation when flows compete.

We retain standard TCP behavior for initialization and timeouts. The rationale is that, at flow start, the sender lacks a meaningful RTT or throughput baseline leaving the LLM without sufficient input to infer a reasonable CWND. In contrast, TCP's ACK-clocked exponential probing of initialization safely discovers usable bandwidth and establishes the context necessary for subsequent decisions.

During a TCP retransmission timeout (RTO), ACKs are absent and the cause of loss is ambiguous, it is thus unclear whether the timeout arises from excessive sending or from a failing link. Unlike a 3-duplicate ACKs, which indicates partial delivery and thus provides more diagnostic information, an RTO offers little guidance. In this case, resetting CWND to one MSS with exponential backoff remains the conservative and well-validated response for preserving stability and fairness; therefore, we retain this behavior in RTO and only incorporate the LLM for 3-duplicate ACKs events. 


\vspace{-2mm}
\subsection{How to Generalize the Triggering Mechanism Across Multiple Phases?}
\label{sec:trigger_generalized}
\vspace{-1mm}

The original latency-based trigger module (\S\ref{sec:trigger_basline}) tends to invoke the LLM prematurely before actual loss, so the LLM repeatedly observed pre-congestion snapshots, usually flat throughput with slight RTT upticks, where the safe response was to hold or decrease the window. This behavior is discussed in details in section (\S\ref{ss:concl}).
We seek to design a triggering module that enables a more comprehensive action space for the LLM, allowing it to increase, decrease, or maintain CWND under more informative network conditions. 

Our initial thought is to periodically trigger the LLM as opposed to adaptively triggering based on latency. We then realize that a fixed time-based cadence would behave inconsistently across links with widely differing link capacities because the same wall-clock interval can span vastly different amounts of delivered data on slow versus fast paths. This would result in oversampling on low-rate links, where frequent triggers add little new information. Hence the LLM can hardly make effective CWND adjustment. On the contrary, it leads to undersampling on high-rate links, where infrequent triggers would not catch fast link dynamics.

To this end, we propose to use the ACK count as the triggering parameter and invoke the LLM whenever the number of ACKs reaches a threshold. 
The threshold is computed using a procedure similar to that of TCP-LLM-L's trigger module. Specifically, we first run TCP-NewReno on the target topology for 10 seconds and record the total number of ACKs received by the sender.  We then round this number to the nearest thousand (for example, in the one-to-one static topology, 7909 ACKs is rounded to 8000 ACKs) and multiply it by a tunable factor $\beta$ to obtain the final threshold. 

As with the estimation-based calculation (\S\ref{sec:trigger_basline}), this threshold can be recomputed if the network conditions are detected changed significantly. By adapting to links of different rates, this ACK-based periodic triggering enables the LLM to operate under a broader range of network conditions and provides greater flexibility in adjusting CWND. Since the generalized ACK-based trigger module does not require the CCA to induce a loss event in order to compute a specific latency, it is natural to ask whether a different CCA can be used for the estimation. After completing the design of the LLM input, we examine this question through an ablation study on the choice of CCA used in the estimation and on the factor $\beta$ (\S\ref{ss:impact_of_beta}). To achieve balancing responsiveness with practicality, we adopt $\beta = 0.1$, corresponding to an ACK threshold of $800$, as the default configuration in our subsequent experiments.

\vspace{-3mm}
\subsection{Rethinking the LLM Prompting}
\vspace{-1mm}

\label{sec:input_design_generalized}
Recall that our initial LLM-input design omits retransmission count because the latency-based trigger prevents actual loss events from surfacing. With the ACK-based trigger, however, packet loss naturally occurs, and the number of retransmitted packets becomes an essential observable. Adding this feature allows the model to better capture congestion events and distinguish them from benign delay variations, thereby enhancing its awareness of network dynamics.

The degree of freedom grants to the LLM also shapes how we design prompts. When the LLM is constrained to operate only during congestion avoidance, as in TCP-LLM-L, hand-crafted rules are required to guide each adjustment. In contrast, when we broaden the model’s scope of intervention, we must rely on more generalized principles because excessive micromanagement risks would overwhelm the model with contradictory directives, while well-structured high-level guidelines give it the flexibility to adapt to diverse network conditions. 
Therefore we propose to encode high-level adjustment principles that anchor the LLM in basic TCP protocol semantics while leaving room for autonomous reasoning, enabling the LLM to leverage its pretrained knowledge rather than merely mimic predefined rules.

\begin{figure*}[t]
  \centering
  \includegraphics[width=\textwidth]{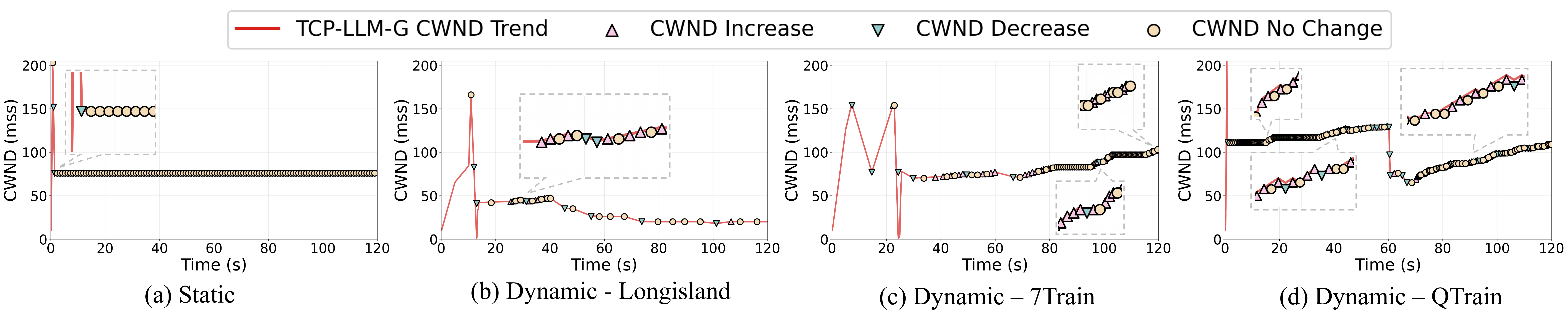}
  \vspace{-7mm}
  \caption{LLM behavior in \textbf{TCP-LLM-G(\S\ref{sec:generalized})} over 120s for 4 topologies: (a) \textbf{Static}, (b) \textbf{Dynamic-Longisland}, (c) \textbf{Dynamic-7Train}, and (d) \textbf{Dynamic-QTrain}.}
  \vspace{-3mm}
  \label{fig:llm_behav_g}
\end{figure*}

We organize the prompt as a progression from core reasoning to generalization and fairness, so that each layer motivates the next and collectively bounds the LLM’s actions. We term this new LLM-based congestion control policy TCP-LLM-G, where G stands for generalization.
Due to page limitation, we put the entire prompt to Appendix Tab.~\ref{tab:gpt_prompts_generalized} while sketching how we design this prompt below.

First, we ground the model in basic AIMD intuition (R1–R4), which defines how congestion and bandwidth underutilization are manifested in the input network statistics and specifies their appropriate handling. This establishes the fundamental control logic. 
Second, to improve the throughput–latency balance (R5), the model should aim to lower queueing without sacrificing goodput. 
Third, we impose protocol-safe bounds (R6–R8), an MSS floor and a prohibition on single-step collapses, to prevent unstable or noncompliant outputs. 
Fourth, we add explicit loss handling (R9): retransmissions serve as a direct congestion signal. 
Fifth, to cope with dynamic bandwidth shifts, the LLM should probe by increasing CWND(R10) but keep this increase only when throughput improves promptly; otherwise, we revert, avoiding optimistic overshoot. 
Finally, to promote long-term fairness (R11–R12), the LLM should temper window growth to avoid crowding out competing flows, and it includes a catch-up mechanism that quickly restores a starved flow to a minimal healthy window to prevent persistent starvation. 

Taken together, this progression gives the LLM meaningful freedom to reason while anchoring its decisions in TCP-safe bounds and measurable performance signals. We term this LLM-based TCP congestion control across multiple congestion control phases as TCP-LLM-G. 

\vspace{-3mm}
\subsection{LLM Behavior Analysis: Generalized Design vs. Limited Design}
\label{ss:deep_dive_generalized}
\vspace{-1mm}

A central question is whether the TCP-LLM-G design provides tangible benefits over TCP-LLM-L in diverse network conditions. To answer this question, we compare these two algorithms both in terms of performance-wise experiment validation and through finer-grained behavior analysis. The following subsections present these results in detail: first by examining their overall performance across topologies, and then by analyzing their CWND control behaviors in response to abrupt bandwidth shifts.

\vspace{-3mm}
\subsubsection{Congestion Control Performance}
\label{sss:metrics_performance}
\vspace{-1mm}

As shown in Fig.~\ref{fig:baseline_limited},  TCP-LLM-G achieves a more favorable trade-off between RTT and throughput compared to TCP-LLM-L. In the highly dynamic QTrain trace, TCP-LLM-G lowers latency by about 12–15\% relative to TCP-LLM-L without hurting the throughput (17. 81 Mbps for TCP-LLM-L vs. 17.89 Mbps for TCP-LLM-G) on a link with an average capacity of about 18 Mbps. This is because unlike TCP-LLM-L which restricts CWND adjustments mainly to no-change or decrease, the generalized design allows the LLM to explore a broader action space. CWND adjustments are permitted not only during congestion avoidance but also in response to loss events, enabling the LLM to propose both conservative and aggressive strategies when appropriate.

Comparing these two approaches' performance across different network conditions (Fig.~\ref{fig:baseline_limited}(a)--(d)), we find that both approaches exhibit nearly identical performance in the static network trace, suggesting that the additional flexibility does not introduce unnecessary variance under stable conditions. In dynamic environments, TCP-LLM-G demonstrates clear advantages. Specifically, in the Longisland trace, throughput improves from 2.83 Mbps to 2.89 Mbps on a link with an average capacity of about 2.9 Mbps, while RTT increases by only ~2\%. This shows that TCP-LLM-G can exploit available bandwidth with little penalty in delay and thereby deliver a superior throughput–latency balance. 

In the more volatile Qtrain trace, where TCP-LLM-L previously struggled, TCP-LLM-G instead markedly improves both throughput and RTT. Across repeated trials, TCP-LLM-G occasionally outperforms Bbr in terms of delay: in roughly 30\% of runs, RTT is reduced from 71.49 ms to around 65.99 ms. This represents a rare case where our design achieves even lower latency than Bbr, while also improving throughput from 17.60 Mbps to 17.91 Mbps on a link with an average capacity of about 18 Mbps. Even outside these best-case runs, the average RTT remains modest at 73.58 ms, showing that the design avoids excessive delay while consistently enhancing throughput, reflecting its adaptability to rapidly changing conditions and its robustness relative to baselines.

\begin{figure*}[t]
    \centering
    \includegraphics[width=\textwidth]{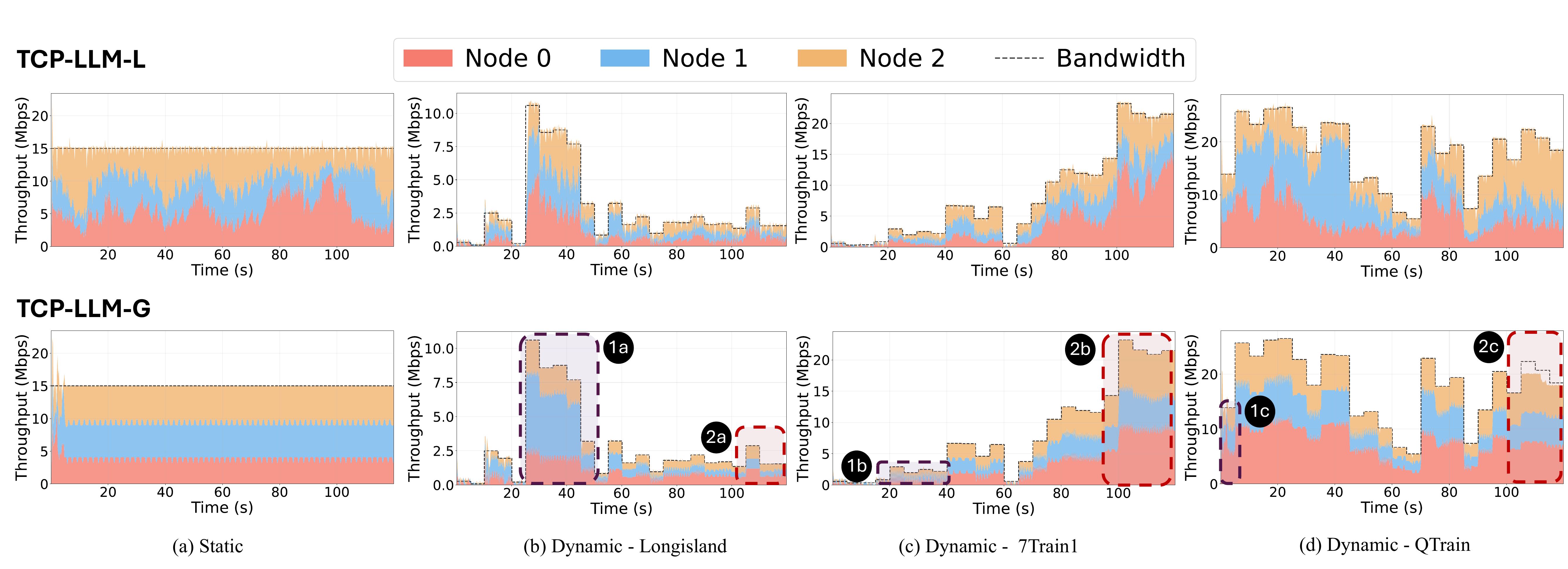}
    \vspace{-8mm}
    \caption{Throughput fairness snapshots over 120 s across four topologies: (a) \textbf{Static}, (b) \textbf{Dynamic-Longisland}, (c) \textbf{Dynamic-7Train}, and (d) \textbf{Dynamic-QTrain} for NewReno, WestwoodPlus, Cubic, Bbr, and our TCP-LLM generalized setup.}
    \label{fig:all_llm_fair}
    \vspace{-3mm}
\end{figure*}

\vspace{-3mm}
\subsubsection{TCP-LLM-G's Fine-grained Behavior Analysis}
\vspace{-1mm}



We further analyze TCP-LLM-G's control behavior on each traces. From Fig.~\ref{fig:llm_behav_g}, a clear behavior is its ability to take the action of increase, which confirms that our tuned prompt effectively expanded the LLM's operational freedom. 

\noindent\textbf{Static.} We observe that the TCP-LLM-G finds a balance point of 76MSS for CWND after four triggers and remains unchanged thereafter in Fig.~\ref{fig:snapshot_final}(a). Notably, Bbr also operates at this value when not probing, suggesting that TCP-LLM-G can interpret the static network condition. 

\noindent\textbf{Longisland.} Turning to the LongIsland trace in Fig.~\ref{fig:snapshot_final}(b), where TCP-LLM-L previously showed an advantage, we find that when bandwidth rises abruptly at 20 seconds, TCP-LLM-G starts pushing the CWND upward for better bandwidth utilization. The specific moderate increases are shown in Fig.~\ref{fig:llm_behav_g}(b), and they enable TCP-LLM-G to quickly adapt when bandwidth drops sharply at 50 seconds. After that, TCP-LLM-G steadily reduces CWND to 20MSS and stays there. This behavior produced a sharply peaked queue size distribution in row 4 of Fig.~\ref{fig:snapshot_final}(a), with a higher concentration at smaller queue sizes, achieving higher throughput while maintaining latency comparable to TCP-LLM-L.

\noindent\textbf{7Train.} Next, Fig.\ref{fig:snapshot_final}(c) shows that when bandwidth ramps up at 60 seconds, TCP-LLM-G also increases CWND to match its observations (Fig.\ref{fig:llm_behav_g}(c)). Compared to TCP-LLM-L, where all increases come from the background TCP-NewReno, TCP-LLM-G’s increases are more conservative. In TCP-LLM-L, the aggressive increases from TCP-NewReno often cause the LLM to respond by decreasing CWND too sharply when triggered (e.g. at 75, 88 seconds), resulting in lower latency but also lower throughput than TCP-LLM-G.

\noindent\textbf{QTrain.} Finally, we switch to the QTrain trace in Fig.~\ref{fig:snapshot_final}(d) and Fig.~\ref{fig:llm_behav_g}(d). Recall that in TCP-LLM-L, the infrequent triggering of the LLM causes it to revert to basic TCP-NewReno and fail to gain advantage over the baselines. In contrast, TCP-LLM-G can recognize the high bandwidth through frequent triggers and increases CWND reasonably to fully utilize the capacity. When the bandwidth drops quickly between 40 seconds and 65 seconds, TCP-LLM-G promptly decreases CWND to respond. During the bandwidth spike from 65 seconds to 85 seconds, we observe an increase at 65 seconds and a decrease at 85 seconds, closely tracking the spike. After 85 seconds, the LLM again increases CWND to match the increasing bandwidth. These timely responses enable TCP-LLM-G to achieve lower latency and higher throughput compared to TCP-LLM-L. This is also demonstrated in row 4 of Fig.~\ref{fig:snapshot_final}(d), where TCP-LLM-G is the only CCA that has its main peak around 20 to 40 packets.

Overall, these analyses show that the generalized action space  with the ACK-based trigger makes TCP-LLM-G more adaptable across diverse topologies, consistently balancing latency and throughput better than TCP-LLM-L.


{\it \noindent\textbf{Takeaway}. Allowing the LLM to act during both congestion avoidance and loss events addresses the limited increases seen in the constrained design. Adding retransmission counts improves visibility into true losses, while an ACK-based trigger exposes broader dynamics than latency alone. These changes let the LLM adapt quickly to fluctuating traces, improving the throughput–latency trade-off, while preserving stability in static conditions.}

%% file: 6fairness.tex
\vspace{-4mm}
\section{Fairness in Multi-Sender Settings}
\label{s:fairness}
\vspace{-3mm}
We further explore whether LLM can achieve fairness.
To this end, we construct a multi-sender–to–one-receiver topology to analyze the third metric: {\it fairness}. We categorize the fairness experiments into two settings: \textbf{All-LLM}, where all three senders employ our LLM-based system for congestion control, and \textbf{Hybrid}, where one sender uses a traditional, widely adopted TCP congestion control algorithm while the remaining two use our system. 

\vspace{-3mm}
\subsection{All-LLM Setups}
\vspace{-1mm}

Fig.~\ref{fig:all_llm_fair} illustrates the per-flow throughput over time. Under TCP-LLM-L, we observe high short-term variability with no clear convergence: flows repeatedly overshoot, trigger losses, and then sharply reduce their CWND. This oscillatory pattern reflects the inherited NewReno-like behavior of the limited design, where aggressive additive increase and multiplicative decrease dominate the dynamics. As a result, individual flows temporarily diverge substantially from one another, even though the long-term averages remain close.

In contrast, the generalized design TCP-LLM-G shows a different trajectory. At the beginning of the trace (labeled Points \textbf{1} in Fig.~\ref{fig:all_llm_fair}), the system is still in its startup phase. Because the LLM has not yet accumulated sufficient information about networks, the flows compete unevenly and the allocations appear unstable. As the transmission progresses, however, the model observes more congestion signals and throughput trends, enabling it to refine its decisions. So, by the later stage of the trace (labeled Points \textbf{2} in Fig.~\ref{fig:all_llm_fair}), the flows converge toward nearly identical throughput, and the allocation stabilizes. This progression indicates that, when granted greater flexibility, the LLM can leverage its generalization ability to learn from the evolving feedback, mitigate oscillations, and guide the system toward a balanced state.


\vspace{-2mm}
\subsection{Hybrid Setups}
\vspace{-2mm}
We next examine fairness in a hybrid setting where nodes employ different CCAs. From our earlier observations, the LLM in TCP-LLM-L consistently produces only non-increasing actions. When competing with a TCP-NewReno flow, which continually increases its CWND until encountering three duplicate ACKs, TCP-LLM-L backs off and is eventually dominated by the NewReno node. Moreover, the deep-dive analyses in Fig.~\ref{fig:snapshot_final} reveal that among the four baseline CCAs we considered, Bbr is the only one without an aggressive increase trend. We therefore focus on the fairness between Bbr and TCP-LLM-G to gain deeper insight.

\begin{figure}[t]
    \centering
    \includegraphics[width=\columnwidth]{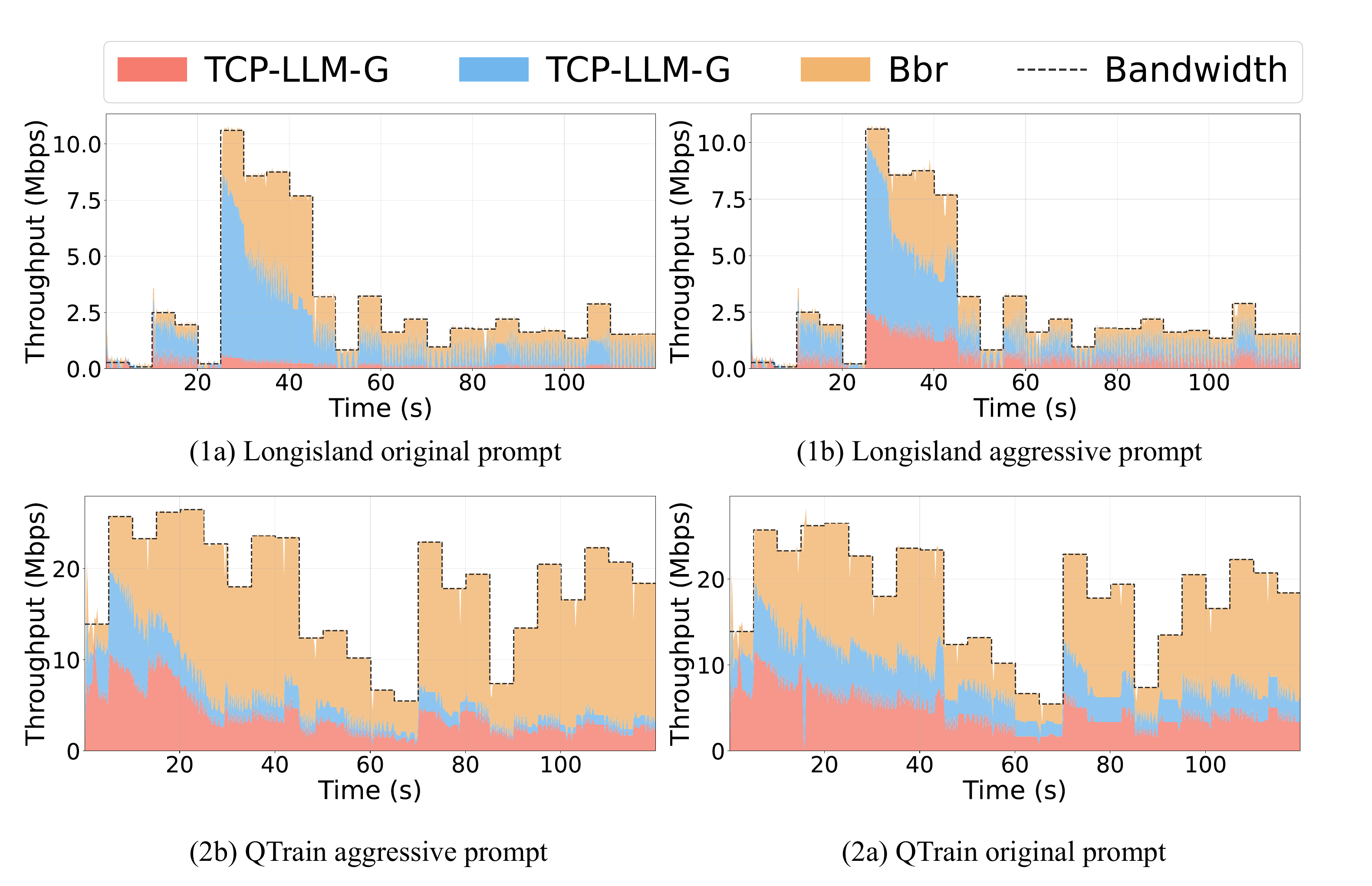}
    \vspace{-6mm}
    \caption{Throughput fairness snapshots over 120s after aggressive prompt.}
    \label{fig:agressive_prompt}
    \vspace{-3mm}
\end{figure}

The `Hybrid' reveals a clear trend: Bbr consistently dominates across all topologies. The TCP-LLM-G flows are relegated to a particularly small share of bandwidth in the Longisland and QTrain topologies as shown in Fig.~\ref{fig:agressive_prompt}(1a)-(1b). To understand the underlying mechanism, we examine the four network statistics features provided to the LLM from the sender’s perspective. We observe that RTT, throughput, and retransmissions exhibit nearly identical patterns under two distinct scenarios: (a) when the available bandwidth decreases, and (b) when a new node joins the network and begins consuming bandwidth. From the LLM's point of view the statistics are: RTT is increasing, throughput is decreasing, and there could be retransmitted packets or not, based on how aggressive the new node is and how large the bandwidth drop is. This similarity makes it difficult for the LLM to distinguish which scenario it is encountering. 

In particular, if it takes a conservative stance and decreases CWND whenever these congestion-like statistics appear, its bandwidth share will inevitably be consumed by more aggressive competitors. This explains why the All-LLM case exhibits greater fairness than the Hybrid case. Conversely, if the LLM attempts to be aggressive and increases CWND whenever it perceives its utilization as too low, it sacrifices its latency advantage in normal conditions. 

To better demonstrate this trade-off, we modify the prompt to instruct the LLM to spike the CWND to probe the network when it considers the current flow to be suppressed by other nodes. We then observe that the average bandwidth utilization of the TCP-LLM-G nodes is much more fair in both the Longisland trace, changing from (9.10\%, 39.10\%, 48.10\%) to (20.50\%, 38.30\%, 37.70\%), and the QTrain trace, changing from (21.50\%, 11.20\%, 66.20\%) to (26.90\%, 21.90\%, 50.20\%). A detailed comparisons are shown in Fig.~\ref{fig:agressive_prompt}, and we can see that the bandwidth utilization for the three nodes are more reasonable after changing the prompt. However, this aggressive change in the prompt causes the average RTT across the three flows to increase from 88.80 ms to 89.54 ms in the QTrain trace, while the difference is minimal in the Longisland trace.

{\it \noindent\textbf{Takeway}. In the all-LLM setting, both TCP-LLM-L and TCP-LLM-G achieve fairness across flows, though TCP-LLM-G converges more smoothly while TCP-LLM-L remains unstable. In hybrid settings, however, the LLM often yields to traditional CCAs due to its less aggressive competition for bandwidth. Adding prompts for stronger increases can restore fairness but at the expense of higher RTT, highlighting a trade-off between equal sharing and latency.}

%% file: 7interpretability.tex
\vspace{-5mm}
\section{Model Interpretability}
\label{s:interpretability}
\vspace{-3mm}

We further translate LLM's congestion control policies into formalized algorithms to better understand their behaviors. 
Specifically, we record sender-side statistics (CWND, RTT, throughput, and retransmissions) together with the LLM’s responses at every intervention point during the QTrain simulations. These logs are then provided back to GPT-5\cite{openai2025gpt5systemcard}, which is asked to summarize the observed behavior and distill it into a rule-based congestion control algorithm. A full set of GPT-5’s generated rules are included in Appendix \ref{tab;gpt_explanation}.

The analysis shows that these control policies converges on a three-rule heuristic: (1) immediate window reduction upon detecting retransmissions, with heavier cuts for persistent or multiple losses; (2) cautious, periodic probing of bandwidth when RTT and throughput appear stable, matching the additive increases we saw in the log; and (3) holding CWND constant when signals are ambiguous, which reflects the plateau behavior in the traces. These generated rules align closely with the actual action patterns exhibited by TCP-LLM in the QTrain, suggesting the model’s decisions are more likely to be interpretable as structured heuristics rather than arbitrary responses. 

Compared to existing CCAs, the LLM-based solution differs in both bandwidth exploration and congestion handling. 

\noindent$\bullet$ In congestion avoidance, unlike NewReno or Cubic that continuously increase CWND until loss signals force a reduction, the LLM only increases CWND after stability is confirmed, gating its probing on the absence of retransmissions, stable RTT, and non-decreasing throughput. This makes probing slower but safer, reducing oscillations in volatile conditions. 

\noindent$\bullet$ For loss handling, unlike Reno and Cubic which always apply a fixed multiplicative decrease, the LLM applies a two-step reduction: a mild cut for small losses and a deeper cut for persistent or heavy losses. Finally, in contrast to Bbr, which aggressively and continuously probes based on its bandwidth–delay model, the LLM remains signal-driven and conservative: it holds CWND  steady when conditions are ambiguous, cutting back only on confirmed loss and increasing only under stability. 

Taken together, the LLM-generated congestion control policies can be seen as a cautious middle ground: less aggressive in growth than Reno, Cubic, and Bbr, but more nuanced in loss handling and more context-aware in its decision logic.

%% file: 8conclusion.tex
\vspace{-5mm}
\section{Conclusion}
\vspace{-3mm}
Our exploratory study with TCP-LLM shows that LLMs can be applied to congestion control in a way that moves beyond fixed rules and adapts to diverse network conditions. While simulations across static and dynamic topologies suggest that TCP-LLM achieves a more favorable throughput–latency trade-off than classic CCAs and offers preliminary insights into fairness, these findings remain early-stage. Important challenges such as computational overhead, stability of LLM decisions, and practical deployment in real-world WAN environments, should be addressed before such approaches can be widely adopted. Looking ahead, we see opportunities in combining LLM reasoning with lightweight traditional control, refining prompts or fine-tuning for efficiency, and extending evaluations to real testbeds. 